\documentclass[vecphys]{svmult}

\usepackage{epsfig}
\usepackage{makeidx}         
\usepackage{graphicx}        
\usepackage{multicol}        
\usepackage[bottom]{footmisc}

\makeindex             

\begin{document}
\title*{The birth of string theory}
\author{Paolo Di Vecchia\inst{1}
}
\institute{Nordita, Blegdamsvej 17, 2100 Copenhagen {\O}, Denmark
\texttt{divecchi@alf.nbi.dk}
}
%
%
\maketitle

\begin{abstract}
In this contribution we go through the developments that in the years from 
1968 to about 1974
led from the Veneziano model to the bosonic string theory. They include the
construction of the $N$-point amplitude for scalar particles, its factorization
through the introduction of an infinite number of oscillators and the proof 
that the physical subspace was a positive definite Hilbert space. We also
discuss  the zero slope limit and the calculation of loop diagrams. Lastly,
we describe how it  finally was recognized that a quantum relativistic string 
theory  was the theory underlying the Veneziano model.
\end{abstract}


\section{Introduction}
\label{intro}

The sixties was a period in which strong interacting processes
were studied in detail using the newly constructed accelerators at Cern
and other places. Many  new hadronic states were found 
that appeared as resonant peaks in various cross sections 
and hadronic cross sections were measured with increasing accuracy.
In general, the experimental data for strongly interacting processes
were rather well understood in terms of
resonance exchanges in the direct channel at low energy and by the
exchange of Regge poles  in the transverse channel at higher energy.  
Field theory that had been very successful in describing  QED
seemed  useless for strong interactions given the big number 
of hadrons to accomodate in a Lagrangian and
the strength of the pion-nucleon coupling constant that did not
allow  perturbative calculations.  The only domain in which field 
theoretical techniques were successfully used was current algebra.
Here, assuming that strong 
interactions were described by an almost chiral invariant
Lagrangian,  that chiral symmetry was 
spontaneously broken and that the pion was the corresponding 
Goldstone boson,   field theoretical methods gave rather 
good predictions for scattering amplitudes involving 
pions at very low energy. Going to higher energy was, however,
not possible with these methods.

Because of this, many people started to think that field theory
was useless to describe strong interactions and tried to 
describe strong interacting processes with alternative 
and more phenomenological methods. The basic ingredients for describing
the experimental data were  at low energy the exchange  of 
resonances in the direct channel and at higher energy 
the exchange of  Regge poles  in the transverse channel. 
Sum rules for strongly 
interacting processes were saturated in this way and one found 
good agreement
with the experimental data that came from the newly constructed
accelerators. Because of these successes and of the problems that
field theory encountered  to  describe the data, it was proposed
to construct directly the S matrix without passing through a 
Lagrangian.  The S matrix was supposed to be constructed 
from the properties that it should satisfy, but there was no clear 
procedure on how to implement this 
construction\footnote{For a discussion of 
S matrix theory see Ref.s \cite{ESSE}}. The word ``bootstrap''
was often used as the way to construct the S matrix, but it did
not help very much to get  an S matrix for the strongly interacting 
processes.

\begin{figure}[ht]
\epsfxsize = 6cm
\centerline{\epsfbox{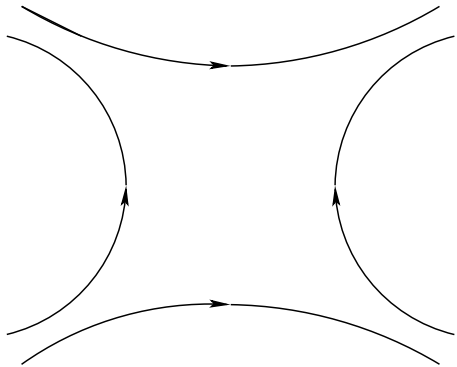}}
\caption{\small Duality diagram for the scattering of four mesons}
\label{duadiagram}
\end{figure}

One of the  basic ideas that led to the construction of an S matrix was that
it should include resonances at low energy and at the same time give 
Regge behaviour at high energy. But the two contributions 
of the resonances and of the Regge poles  should not
be added because this would imply double counting.
This was called  Dolen, Horn and  Schmidt 
duality~\cite{DHS}.  Another idea that
helped in the construction of an 
S matrix was  planar duality~\cite{HR}  that was 
visualized by associating to a certain process a duality 
diagram, shown in Fig. (\ref{duadiagram}), where each meson was described by  
two lines representing the quark and the antiquark. Finally, also the 
requirement of crossing symmetry played a very important role.

Starting from these ideas  Veneziano~\cite{VENE} 
was able to construct an S matrix for the
scattering of four mesons that, at the same time, had 
 an infinite number of zero width 
resonances lying on linearly rising Regge trajectories and  
 Regge behaviour at high energy.
Veneziano originally constructed  the model for the process 
$\pi \pi \rightarrow \pi \omega$, but it was immediately extended to
the scattering of four scalar particles.

\begin{figure}[ht]
\epsfxsize = 6cm
\centerline{\epsfbox{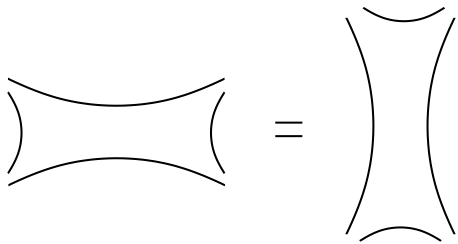}}
\caption{\small The duality diagram contains both s and t channel poles}
\label{dua}
\end{figure}

In the case of four identical scalar particles, the
crossing symmetric scattering amplitude found by Veneziano consists of
a sum of three terms:
\begin{eqnarray}
A (s, t, u) = A(s,t) + A(s, u) + A(t,u)
\label{astu}
\end{eqnarray}
where
\begin{eqnarray}
A(s,t) = \frac{\Gamma (- \alpha (s) )\Gamma (- \alpha (t) )}{
\Gamma (- \alpha (s) - \alpha (t) )} = \int_{0}^{1} dx x^{- \alpha(s)
-1} (1-x)^{-\alpha (t) -1}
\label{ast}
\end{eqnarray}
with linearly rising Regge trajectories
\begin{eqnarray}
\label{rising}
\alpha{(s)} = \alpha_{0} + \alpha ' s
\end{eqnarray}
This was a very important property to implement 
in a model  because it was in
agreement with the experimental data in a wide range of energies.
$s,t$ and $u$ are the Mandelstam variables:
\begin{eqnarray}  
\label{stu}
s= -(p_{1}+ p_{2})^{2} \hspace{.5cm},\hspace{.5cm}
t= -(p_{3}+ p_{2})^{2} \hspace{.5cm},\hspace{.5cm}
u= -(p_{1}+ p_{3})^{2}
\end{eqnarray}
The three terms in Eq. (\ref{astu}) correspond to the three
orderings  of the four particles that are not related by a cyclic or
anticyclic~\footnote{An anticyclic permutation corresponding, for 
instance, to the ordering $(1234)$ is obtained by taking 
the reverse of the original ordering $(4321)$  and then performing a cyclic
permutation.}  permutation of the external legs. They 
correspond, respectively, to the three permutations: $(1234), (1243)$
and $(1324)$ of the four external legs. 
They have only simple pole singularities. The first one 
has only poles in the s and t channels, the second 
only in the s and u channels and the third only in the t and u
channels. This property follows directly from the duality diagram that
is associated to each inequivalent permutation of the external legs. 
In fact, at that time one used to associate to each of the three
inequivalent  permutations a duality diagram where each particle was
drawn as consisting of two lines that rappresented the quark and
antiquark making up a meson. Furthermore, the diagram was supposed to
have only poles singularities in the planar channels which are those
involving adjacent external lines. This means that, for instance, the
duality diagram corresponding to the permutation $(1234)$ has only
poles in the s and t channels as one can see by deforming the diagram 
in the plane in the two possible ways shown in figure (\ref{dua}).
 
This 
was a very important property of the duality diagram  that makes it
qualitatively different from a Feynman diagram in
field theory  where each diagram has only a pole in one of the three
s, t and u channels and not simultaneously in two of them. If we
accept the idea that each term of the sum in Eq. (\ref{astu}) is
described by a duality diagram, then it is clear that we do not need to
add terms corresponding to equivalent diagrams because the
corresponding duality diagram is the same and has the same singularities.
It is now clear that it was in some way implicit in this picture the fact 
that  the Veneziano model  corresponds  to the scattering 
of relativistic strings. But at that time the connection 
was not obvious at all.
The only S matrix property that the Veneziano model failed 
to satisfy was the unitarity of the S matrix. because it contained only
zero width resonances and did not have the various  cuts
required by unitarity. We will see how this property will be  implemented.

Immediately after the formulation of the Veneziano model,
Virasoro~\cite{VIRA} proposed another 
crossing symmetric four-point amplitude  for scalar particles that
consisted of a unique piece given by:
\begin{eqnarray}
A (s, t , u)  \sim  \frac{\Gamma ( - \frac{\alpha (u)}{2})   \Gamma 
( - \frac{\alpha (s)}{2}) \Gamma (- \frac{\alpha (t)}{2}) }{ 
\Gamma ( 1 + \frac{\alpha (u)}{2}) \Gamma ( 1 + \frac{\alpha (s)}{2})
\Gamma ( 1 + \frac{\alpha (t)}{2}) }
\label{vir}
\end{eqnarray}
where
\begin{eqnarray}
\alpha (s) = \alpha_0 + \alpha' s
\label{tra39}
\end{eqnarray}
The model had poles in all three $s, t$ and $u$ channels and
could not be written as sum of three terms having poles only in planar
diagrams.  In conclusion, the Veneziano model satisfies the principle 
of planar duality being a crossing symmetric combination of three contributions
each having  poles only in the planar channels. On the other hand, the 
Virasoro model consists of a unique crossing symmetric term having poles
in both planar and non-planar channels.

The attempts to construct consistent  models that were in good agreement with
the strong interaction phenomenology of the sixties  boosted enormously
the activity in this research field. The generalization of the Veneziano model 
to the scattering of $N$  scalar particles  
was built, an operator formalism consisting of
an infinite number of harmonic oscillators was constructed
and the complete spectrum of mesons was
determined. It turned out that the degeneracy of states grew up 
exponentially with the mass.  It was also found that the  $N$ point amplitude 
had states with negative norm (ghosts) unless the intercept of the
Regge trajectory was $\alpha_0 =1$~\cite{VIRA2}. 
In this case it turned out that the model was free of
ghosts but the lowest state was a tachyon.  The model was called
in the literature the ``dual resonance model''.

The model was not unitary because all the states were zero width
resonances and the various cuts required by unitarity were absent.
The unitarity was implemented in a perturbative way 
by adding loop diagrams 
obtained by sewing some of the external legs together 
after the insertion of a propagator.  The multiloop
amplitudes showed  a structure of Riemann surfaces. This became 
obvious only later  when the dual resonance model was recognized
to correspond to scattering of strings.

But the main problem was that the model had a tachyon if 
$\alpha_0 =1$ or had ghosts for other values of $\alpha_0$ and 
was not in agreement with the experimental data: $\alpha_0$ 
was  not equal to  about $\frac{1}{2}$ as required by experiments for 
the $\rho$ Regge trajectory and the external scalar particles
did not behave as pions satisfying the current algebra 
requirements.  Many attempts were made to construct more
realistic dual resonance models, but the main result of these attempts
was the construction of the Neveu-Schwarz~\cite{NS} and  
the Ramond~\cite{RAMOND}  models,
respectively, for mesons and fermions. They were constructed as two independent
models and only later  were recognized to be two sectors of
the same model.  The Neveu-Schwarz model still contained a tachyon 
that only in 1976 through the GSO projection was eliminated from
the physical spectrum. Furthermore,  it  was not properly 
describing the properties
of the physical pions.

Actually a model describing $\pi \pi$ scattering in a rather satisfactory way  
was proposed by Lovelace and Shapiro~\cite{LOVE}~\footnote{See 
also Ref.~\cite{FRAMPTON}.}. According to this
model the three isospin amplitudes for pion-pion scattering are given by:
\[
A^0 = \frac{3}{2} \left[ A(s,t) + A (s,u) \right] - \frac{1}{2} A(t,u)
\]
\begin{eqnarray}
A^1 = A(s,t) - A (s,u) \hspace{2cm} A^2 = A (t,u)
\label{isoamp}
\end{eqnarray}
where
\begin{eqnarray}
A(s,t) = \beta \frac{\Gamma (1 - \alpha (s) ) 
\Gamma (1 - \alpha (t))}{\Gamma (1 - \alpha (t) - \alpha (s) )}
\hspace{1cm};\hspace{1cm} \alpha (s) = \alpha_0 + \alpha' s
\label{astbis}
\end{eqnarray}
The amplitudes in eq.(\ref{isoamp}) provide
a model for $\pi \pi$ scattering with linearly rising Regge trajectories 
containing three parameters: the intercept of the $\rho$ Regge trajectory 
$\alpha_0$, the Regge slope $\alpha'$ and $\beta$. The first two can be
determined by imposing the Adler's self-consistency condition, that requires
the vanishing of the amplitude when $s=t=u = m_{\pi}^{2}$ and one of the pions
is massless, and  the fact that the Regge trajectory must give the 
spin of the $\rho$ meson that is equal to $1$ when $\sqrt{s}$ is equal to the
mass of the $\rho$ meson $m_{\rho}$. These two conditions determine the Regge
trajectory to be:
\begin{eqnarray}
\alpha (s) = \frac{1}{2} \left[1 + 
\frac{s - m_{\pi}^{2}}{m_{\rho}^{2} - m_{\pi^{2}}} \right] = 0.48 + 0.885 s
\label{rhotra}
\end{eqnarray}
Having fixed the parameters of the Regge trajectory the model predicts 
the masses and the couplings of the resonances that decay in $\pi \pi$ 
in terms of a unique parameter $\beta$. The values obtained are in 
reasonable agreement with the experiments. Moreover,  one can compute the
$\pi \pi$ scattering lenghts:
\begin{eqnarray}
a_0 = 0.395 \beta \hspace{2cm} a_2 = - 0.103 \beta
\label{scatle}
\end{eqnarray}
and one finds that their
 ratio is within $10 \%$ of the current algebra ratio given by $a_0 /a_2
= - 7/2$. 
The amplitude in eq.(\ref{astbis}) has exactly the same form as that for
 four tachyons 
 of the Neveu-Schwarz model with the only apparently minor difference
that $\alpha_0 =1/2$ (for $m_{\pi} =0$) instead of $1$ as in the Neveu-Schwarz 
model. This difference, however, implies that the critical space-time 
dimension of this
model is $d=4$~\footnote{This can be checked by computing the coupling of the 
spinless particle at the level $\alpha (s) =2$ and seeing that it vanishes for
$d=4$.} and not  $d=10$ as in the Neveu-Schwarz model. 
In conclusion this model seems to 
be a
perfectly reasonable model for describing low-energy $\pi \pi$ scattering. The
problem is, however, that nobody has been able to generalize it to the 
multipion scattering and therefore to get the complete meson spectrum.

As we have seen the S matrix of the dual resonance model was 
constructed using ideas and tools of hadron phenomenology of the
end of the sixties. Although it did not seem 
possible  to write a realistic 
dual resonance model describing the pions ,  it was nevertheless 
such a source of fascination for  those who actively worked 
in this field at that time for its beautiful internal structure 
and consistency that a lot of energy was used to investigate its 
properties and for understanding its basic structure.
It turned out with great surprise that the underlying structure
was that of a quantum relativistic string. 

The aim of this contribution is to explain the logic of  the work that was 
done in the years from 1968 to 1974~\footnote{Reviews from this period
  can be found in Ref.~\cite{REV}} in order to uncover the deep 
properties of this  model   that appeared from the beginning to be so
beautiful and consistent to deserve  an intensive study. 

This seems to me a very good way of celebrating the 65th anniversary
of Gabriele who is the person who started and also contributed to 
develop the whole thing with his deep physical intuition.

\section{Construction of the $N$-point amplitude}
\label{npoint}

We have seen that the construction of the four-point amplitude is not
sufficient  to get information on the full hadronic spectrum because it
contains only those hadrons that couple to two ground state mesons
and does not see those intermediate states which only couple to three or
to an higher number of ground state 
mesons~\cite{GLIO}. Therefore, it was very important to construct the $N$-point
amplitude involving identical scalar particles. The construction
of the $N$-point amplitude was done in Ref.~\cite{NPARTI} (extending the 
work of Ref.~\cite{5POINT}) by 
requiring the same
principles that have led to the construction of the
Veneziano model, namely the fact
that the axioms of S-matrix theory be satisfied by an infinite number
of zero width resonances lying on linearly rising Regge trajectories
and planar duality. 

The fully crossing symmetric scattering amplitude of $N$ identical
scalar particles is given by a sum of terms corresponding to the
inequivalent permutations of the external legs:
\begin{eqnarray}
A = \sum_{n=1}^{N_p} A_{n} 
\label{across}
\end{eqnarray}
Also in this case two permutations of the external legs are
inequivalent if they are not related by a cyclic or anticyclic permutation.
$N_p$ is the number of inequivalent permutations of the external legs
 and is equal to  $N_p =
\frac{(N-1)!}{2}$ and each term has only simple pole singularities in the
planar channels. Each planar channel is described by two indices
$(i,j)$, to mean that it includes the legs $i, i+1, i+2 \dots j-1, j$, 
by the Mandelstam variable
\begin{eqnarray}
s_{ij} = - (p_i + p_{i+1} + \dots + p_j )^2
\label{sij}
\end{eqnarray}
and by an additional variable $u_{ij}$ whose role will become clear soon.
It is clear that the channels $(ij)$ and 
$(j+1 , i-1)$~\footnote{This channel includes the 
particles  $(j+1, \dots , N, 1, \dots i-1)$.}  are identical
and they should be counted only once.
In the case of N identical scalar particles the number of planar
channels is equal to $\frac{N(N-3)}{2}$. This can be obtained as
follows. The independent planar diagrams involving the
particle 1 are of the type $(1,i)$ where $i=2 \dots N-2$. Their number
is $N-3$. This is also the number of planar diagrams involving the
particle 2 and not the 1. The number of planar diagrams involving the
particle 3 and not the particles 1 and 2 is equal to $N-4$. In
general the number of planar diagrams involving the particle i and not
the previous ones from 1 to i-1 is equal to $N-1-i$. This means that
the total number of planar diagram is equal to:
\[
2 (N-3) + \sum_{i=3}^{N-2} (N-1-i) = 2 (N-3) + \sum_{i=1}^{N-4} i =
\]
\begin{eqnarray}
= 2 (N-3) + \frac{(N-4)(N-3)}{2} = \frac{N(N-3)}{2}
\label{planardi}
\end{eqnarray}
If one writes down the
duality diagram corresponding to a certain planar ordering of the
external particles, it is easy to see that the diagram can have
simultaneous pole singularities only in $N-3$ channels. The channels
that allow simultaneous pole singularities are called compatible
channels, the other are called incompatible. Two channels (i,j) and
(h,k)  are incompatible if the following inequalities are satisfied:
\begin{eqnarray}
i \leq h \leq j~~~;~~~j+1 \leq k \leq i-1
\label{incopa}
\end{eqnarray}
The aim is to construct the scattering amplitude for each inequivalent
permutation of the external legs that has only pole singularities in
the $\frac{N(N-3)}{2}$ planar channels. We have also to impose 
that the amplitude has 
simultaneous poles only in $N-3$ compatible channels. In order to gain
intuition on how to proceed we rewrite the four-point amplitude in
Eq. (\ref{ast}) as follows:
\begin{eqnarray}
A (s, t) = \int_{0}^{1} d u_{12} \int_{0}^{1} d u_{23} \,\, u_{12}^{-
  \alpha (s_{12}) -1}  u_{23}^{- \alpha (s_{23}) -1} \delta (
  u_{12} + u_{23} -1 ) 
\label{ast1}
\end{eqnarray}
where $u_{12}$ and $u_{23}$ are the variables corresponding to the
two planar channels  $(12)$ and $(23)$ and the cancellation 
of simultaneous poles in
incompatible channels is provided by the $\delta$-function which
forbids   $u_{12}$ and $u_{23}$ to  vanish simultaneously. 

We will now extend this procedure  to
the $N$-point amplitude. But for the sake of clarity 
let us start with the case of $N=5$~\cite{5POINT}. 
In this case we have 5 planar channels
described by $u_{12} , u_{13}, u_{23} , u_{24}$ and $ u_{34}$. Since
we have only two compatible channels only two of the previous five
variables are independent. We can choose them to be $u_{12}$ and
$u_{13}$. In order to determine the dependence of the other three
variables on the two independent ones, we exclude simultaneous poles in
incompatible channels. This can be done by imposing relations that
prevent  variables corresponding to incompatible channels to vanish
simultaneously. A sufficient condition for excluding simultaneous
poles in incompatible channels is to impose the conditions:
\begin{eqnarray}
u_{P} = 1 - \prod_{\bar{P}} u_{ {\bar{P}}}
\label{exclu}
\end{eqnarray}
where the product is over the variables ${\bar{P}}$ corresponding to 
channels that are incompatible with $P$. In the case of the five-point 
amplitude we get the following relations:
\[
u_{23} = 1 - u_{34} u_{12}~~;~~u_{24} = 1 - u_{13} u_{12}
\]
\begin{eqnarray}
u_{13} = 1 - u_{34} u_{24}
~~;~~u_{34} = 1 - u_{23} u_{13}~~;~~
u_{12} = 1 - u_{24} u_{23}
\label{rel89}
\end{eqnarray}
Solving them in terms of the two independent ones we get:
\begin{eqnarray}
u_{23} = \frac{1 - u_{12}}{1 - u_{12} u_{13}}~~;~~
u_{34} = \frac{1 - u_{13}}{1 - u_{12} u_{13}}~~;~~ u_{24} = 1 -
u_{12}u_{13} 
\label{sol67}
\end{eqnarray}
In analogy with what we have done for the four-point amplitude in
Eq. (\ref{ast1}) we write the five-point amplitude as follows:
\[
\int_{0}^{1} d u_{12} \int_{0}^{1} d u_{13} \int_{0}^{1} d u_{23} 
\int_{0}^{1} d u_{24} \int_{0}^{1} d u_{34}
u_{12}^{- \alpha (s_{12}) -1} u_{13}^{- \alpha (s_{13}) -1}  \times
\]
\[
\times u_{24}^{- \alpha (s_{24}) -1} 
u_{23}^{ - \alpha (s_{23}) -1} u_{34}^{- \alpha (s_{34}) -1} \times   
\]
\begin{eqnarray}
\delta ( u_{23} + u_{12} u_{34}-1  ) \delta ( u_{24} + u_{12} u_{13}
-1 )
\delta ( u_{34} + u_{13} u_{23} -1 )  
\label{fiv}
\end{eqnarray}
Performing the integral over the variables $u_{23}, u_{24}$ and $
u_{34}$ we get:
\[ 
\int_{0}^{1} d u_{12} \int_{0}^{1} d u_{13}
u_{12}^{- \alpha (s_{12}) -1} u_{13}^{- \alpha (s_{13}) -1} \times
\]
\begin{eqnarray}
\times (1 - u_{12})^{ - \alpha (s_{23}) -1} (1 - u_{13})^{ - \alpha (s_{13})
  -1}
(1 - u_{12} u_{13} )^{- \alpha ( s_{24}) + \alpha ( s_{23}) +  
\alpha ( s_{34})}
\label{fivb}
\end{eqnarray}
We have implicitly assumed that the  Regge trajectory is the same in all
channels and that the external scalar particles have the same
common mass $m$ and are the lowest lying states on the Regge trajectory. 
This means that their mass  is given by:
\begin{eqnarray}
\alpha_0 - \alpha' p_{i}^2 =0~~;~~p_{i}^{2}  \equiv  - m^2
\label{m2}
\end{eqnarray}
Using then the relation: 
\begin{eqnarray}
\alpha (s_{23} )  +  \alpha (s_{34} ) - \alpha (s_{24} ) = 2 \alpha' p_2
\cdot p_4
\label{rel72}
\end{eqnarray}
we can rewrite Eq. (\ref{fivb}) as follows:
\[ 
B_5 = \int_{0}^{1} d u_2 \int_{0}^{1} d u_3  u_{2}^{- \alpha (s_2 ) -1} 
u_{3}^{- \alpha (s_3 ) -1} (1- u_2 )^{-\alpha ( s_{23} ) -1}  \times
\]
\begin{eqnarray}
\times (1-u_3)^{- \alpha ( s_{34} )  -1}
\prod_{i=2}^{2}  \prod_{j=4}^{4} ( 1 - x_{ij} )^{2 \alpha' p_i \cdot p_j} 
\label{b5}
\end{eqnarray}
where
\begin{eqnarray}
s_i \equiv s_{1i}~~,~~u_i \equiv u_{1i}~;~i=2,3 ~~;~~x_{ij} = u_i u_{i+1}
\dots u_{j -1}.
\label{con62}
\end{eqnarray}
We are now ready to construct the $N$-point function~\cite{NPARTI}. 
In analogy with 
what has been done for the four and five-point amplitudes we can write 
the $N$-point amplitude as follows:  
\begin{eqnarray}
B_N =  \int_{0}^{1}  \dots   \int_{0}^{1}  \prod_P  [ 
u_{P}^{- \alpha (s_P  ) -1}  ]
\prod_{Q} \delta ( u_Q -1 + \prod_{\bar{Q}} u_{{\bar{Q}}}    )  
\label{bn}
\end{eqnarray}
where the first product is over the  $\frac{N(N-3)}{2}$ variables 
corresponding to all planar channels, while the second one is 
over  the $\frac{(N-3)(N-2)}{2}$ independent $\delta$-functions. 
The product in the $\delta$-function is 
defined in Eq. (\ref{exclu}).

The solution of all the non-independent linear relations imposed by the 
$\delta$-functions   is given by
\begin{eqnarray}
u_{ij} = \frac{ (1- x_{ij} ) (1 - x_{i-1,j+1}) }{ (1- x_{i-1,j})(1- x_{i,j+1})}
\label{uij} 
\end{eqnarray}
where the variables $x_{ij}$ are given in Eq. (\ref{con62}). Eliminating the 
$\delta$-function from Eq. (\ref{bn}) one gets:
\begin{eqnarray}
B_N = \prod_{i=2}^{N-2} \left[ \int_{0}^{1} d u_i u_{i}^{- \alpha ( s_i
  ) -1} (1 - u_i )^{- \alpha ( s_{i, i+1} ) -1 } \right] \prod_{i=2}^{N-3}
  \prod_{j=i+2}^{N-1} ( 1 - x_{ij} )^{- \gamma_{ij}}
\label{bna}
\end{eqnarray}
where
\begin{eqnarray}
\gamma_{ij} = \alpha ( s_{ij} )  +\alpha ( s_{i+1;j-1} ) - \alpha (
s_{i;j-1}) -\alpha ( s_{i+1;j} ) ~~~;~~ j \geq i+2 
\label{gaij}
\end{eqnarray}
It is easy to see that 
\begin{eqnarray}
\alpha ( s_{i, i+1} )= - \alpha_0 - 2 \alpha' p_i \cdot p_{i+1}~~;~~
\gamma_{ij} = - 2 \alpha' p_i \cdot p_j ~~;~~ j \geq i+2 
\label{su}
\end{eqnarray}
Inserting them in Eq. (\ref{bna}) we get:
\begin{eqnarray}
B_N = \prod_{i=2}^{N-2} \left[ \int_{0}^{1} d u_i u_{i}^{- \alpha ( s_i
  ) -1} (1 - u_i )^{\alpha_0 -1} \right] \prod_{i=2}^{N-2} 
\prod_{j = i+1}^{N-1} (1
  - x_{ij} )^{2 \alpha' p_i \cdot p_j }
\label{bnb}
\end{eqnarray}
This is the form of the $N$-point amplitude that was originally
constructed. Then Koba and Nielsen~\cite{KN} put it in the form that is more
known nowadays. They constructed it using the following rules.
They associated a real variable  $z_i$ to each  leg $i$.
Then they associated to each channel $(i,j)$ an anharmonic ratio  constructed
    from the variables $z_i , z_{i-1} , z_j , z_{j+1}$ in the following way
\begin{eqnarray}
( z_i , z_{i+1} , z_j , z_{j+1} )^{- \alpha ( s_{ij} )-1}  = \left[
\frac{( z_i - z_j) ( z_{i-1} - z_{j+1} ) }{( z_{i-1} - z_{j}) 
( z_{i} - z_{j+1} ) }
\right]^{- \alpha ( s_{ij} )-1 }
\label{anah6}
\end{eqnarray}
and finally they gave the following expression for the $N$-point amplitude:
\begin{eqnarray}
B_N = \int_{- \infty}^{\infty} d V (z) \prod_{(i ,j)}  
( z_i , z_{i+1} , z_j , z_{j+1} )^{- \alpha ( s_{ij} )-1} 
\label{bnc}
\end{eqnarray}
where
\begin{eqnarray}
 d V (z) = \frac{ \prod_{1}^{N} \left[\theta (z_i - z_{i+1}) d z_i
   \right]}{\prod_{i=1}^{N} ( z_{i} - z_{i+2}
 ) d V_{abc}}~;~ d V_{abc} = \frac{dz_a dz_b dz_c}{(z_b -z_a
 )(z_c - z_b) (z_a - z_c)} 
\label{dv}
\end{eqnarray}
and the variables $z_i$ are integrated along the real axis in a cyclically
ordered way: $z_1 \geq z_2 \dots \geq z_N$ with $a, b, c$ arbitrarily
chosen.
 
The integrand of the $N$-point amplitude is invariant under projective
transformations acting on the leg variables $z_i$:
\begin{eqnarray}
z_i \rightarrow \frac{\alpha  z_i + \beta}{\gamma  z_i +\delta }~~;~~
i=1 \dots N~~;~~\alpha \delta - \beta \gamma =1
\label{proj}
\end{eqnarray}
This is because both the anharmonic ratio in Eq. (\ref{anah6}) 
 and the measure $dV_{abc}$ are invariant under a projective transformation. 
Since a projective
transformation depends on three real parameters, then the integrand of
the $N$-point amplitude depends only on $N-3$ variables $z_i$. In
order to avoid infinities, one has then to divide the integration
volume with the factor  $dV_{abc}$ that is also invariant under the
projective transformations. The fact that the integrand depends only
on $N-3$ variables is in agreement with the fact that $N-3$ is also
the maximal number of simultaneous poles allowed in the amplitude. 

It is convenient to write the $N$-point amplitude in a form that
involves the scalar product of the external momenta rather than the
Regge trajectories. We distinguish three kinds of channels. The first one
is when  the particles $i$ and $j$ of the channel $(i,j)$ are
separated by at least two particles. In this case the channels that
contribute to the exponent of the factor $(z_i -z_j)$ are the
channels $(i,j)$ with exponent equal to $- \alpha (s_{ij} ) -1$, $(i+1,
j-1 )$ with exponent $- \alpha ( s_{i+1 , j-1 }) -1$, $(i+1 , j)$ with
exponent $ \alpha ( s_{i+1 , j}) +1$ and $(i, j-1)$ with exponent $ 
\alpha ( s_{i, j-1}) + 1$. Adding these four contributions one gets
for the channels where $i$ and $j$ are separated by at least two
particles 
\begin{eqnarray}
- \alpha (s_{ij} ) - \alpha ( s_{i+1 , j-1 }) + \alpha ( s_{i+1 , j})+
 \alpha ( s_{i, j-1}) = 2 \alpha ' p_i \cdot p_j
\label{one}
\end{eqnarray}
The second one comes from the channels that are separated by only one
particle. In this case only three of the previous four channels
contribute. For instance if $j=i+2$ the channel $(i+1 , j-1)$ consists
of only one particle and therefore should not be included. This means
that we would get:
\begin{eqnarray}
-\alpha (s_{i;i+2}) -1 +\alpha ( s_{1+1; i+2}) +1 +\alpha ( s_{i; i+1}  +1
 ) = 1 + 2 \alpha ' p_{i} \cdot p_{i+2}
\label{two}
\end{eqnarray}
Finally the third one that comes from the channels whose particles are 
adjacent, gets
only contribution from:
\begin{eqnarray}
- \alpha ( s_{i;i+1} ) -1 = \alpha_0 -1 + 2 \alpha ' p_{i} \cdot p_{i+1}
\label{three}
\end{eqnarray}
Putting all these three terms together in Eq. (\ref{bnc}) and
remembering  the factor in the denominator  in the first equation of  
(\ref{dv})
we get:
\begin{eqnarray}
B_N = \int_{-\infty}^{\infty} \frac{\prod_{1}^{N} d z_i  
\theta (z_i - z_{i+1}) }{dV_{abc}}
\prod_{i=1}^{N} \left[ ( z_{i} - z_{i+1} )^{\alpha_0 -1} \right] 
\prod_{j>i} ( z_i -
z_j )^{2 \alpha' p_i \cdot p_j}
\label{bnd}
\end{eqnarray}
A convenient choice for the three variables to keep fixed is:
\begin{eqnarray}
z_a = z_1 = \infty~~;~~ z_b = z_2 = 1~~;~~ z_c = z_N =0
\label{cho8}
\end{eqnarray}
With this choice the previous equation becomes:
\[
B_N = \prod_{i=3}^{N-1}\left[ \int_{0}^{1} d z_i \theta (z_i - z_{i+1}
  ) \right] \prod_{i=2}^{N-1} ( z_i - z_{i+1} )^{\alpha_0 -1} \times
\] 
\begin{eqnarray}  
\times  
\prod_{i=2}^{N-1} \prod_{j=i+1}^{N} (z_i - z_j )^{2 \alpha' p_i \cdot p_j} 
\label{bne}
\end{eqnarray}
We now want  to show that this  amplitude  is identical to the one given in 
Eq. (\ref{bnb}). This can be done by performing the following change 
of variables:
\begin{eqnarray}
u_i = \frac{z_{i+1}}{z_i}~~;~~i= 2, 3 \dots N-2 
\label{new3}
\end{eqnarray}
that implies
\begin{eqnarray}
z_{i} = u_{2}  u_{3} \dots u_{i-1}~~;~~i= 3, 4 \dots N-1
\label{new4}
\end{eqnarray}
Taking into account that the Jacobian is equal to:
\begin{eqnarray}
\det \frac{\partial  z}{\partial u} = 
\prod_{i=3}^{N-2} z_i = \prod_{i=2}^{N-3} 
u_{i}^{N-2 -i}  
\label{det56}
\end{eqnarray}
using the following two relations:
\begin{eqnarray}
\det \frac{\partial  z}{\partial u} 
\prod_{i=2}^{N-1} ( z_i - z_{i+1} )^{\alpha_0 -1} =
 \prod_{i=2}^{N-2} \left[ u_{i}^{(N -1 -i)\alpha_0 -1} \right]   
\prod_{i=2}^{N-2} (1 - u_{i} )^{\alpha_0 -1}
\label{ii+1}
\end{eqnarray}
and
\[
\prod_{i=2}^{N-1} \prod_{j=i+1}^{N} (z_j - z_i )^{2 \alpha' p_i \cdot p_j} 
= 
\]
\begin{eqnarray}
= \prod_{i=2}^{N-2} \prod_{j=i+1}^{N-1}
 ( 1 - x_{ij})^{2 \alpha' p_i \cdot p_j} \prod_{i=2}^{N-2} u_{i}^{ -
   \alpha ( s_{i} ) - (N-i -1) \alpha_0} 
\label{ij}
\end{eqnarray}
and the conservation of momentum
\begin{eqnarray}
\sum_{i=1}^{N} p_i =0
\label{cons}
\end{eqnarray}
together with Eq. (\ref{m2}), one can easily see that Eq.s (\ref{bnb}) and
(\ref{bne}) are equal. 

The $N$-point amplitude that we have constructed in this section corresponds
to the scattering of  $N$ spinless particles with 
no internal degrees of freedom.
On the other hand it was known that the mesons were classified according to
multiplets of an $SU(3)$ flavour symmetry. This was implemented by 
Chan and Paton~\cite{CP}  by multiplying the $N$-point amplitude 
with a factor, 
called Chan-Paton factor, given by
\begin{eqnarray}
Tr ( \lambda^{a_1} \lambda^{a_2} \dots \lambda^{a_N} ) 
\label{chanpa}
\end{eqnarray}
where the $\lambda$'s  are  matrices of a unitary group in the 
fundamental representation.  Including the Chan-Paton 
factors the total scattering amplitude is given by:
\begin{eqnarray}
\sum_{P}  Tr ( \lambda^{a_1} \lambda^{a_2} \dots \lambda^{a_N} )  
B_{N} (p_1 , p_2 , \dots p_N )
\label{chapa}
\end{eqnarray}
where the sum is extended to the $(N-1)!$ permutations of the 
external legs, that are not related by a cyclic permutations. 
Originally  when the dual resonance 
model was supposed
to describe strongly interacting mesons, this factor was introduced 
to represent their 
flavour degrees of freedom. Nowadays the interpretation is different 
and the Chan-Paton factor represents the colour degrees of freedom of the
gauge bosons and the other massive particles of the spectrum.

The $N$-point amplitude $B_N$ that we have constructed in this section 
contains only simple pole singularities in all possible planar
channels. They correspond to zero width resonances located at  
non-negative integer values $n$ of the Regge trajectory 
$\alpha ( M^2 ) = n$. The lowest state located at $\alpha ( m^2) =0$ 
corresponds 
to the particles on the external legs of $B_N$. The  spectrum of 
excited particles can be obtained by factorizing the $N$-point 
amplitude in the most general channel with any number of particles. 
This was done in Ref.s ~\cite{FUVE1} and  \cite{BAMA} finding a 
spectrum of states rising exponentially with the mass $M$.  Being the 
model relativistic invariant  it was found that many states obtained
by factorizing the $N$-point amplitude were "ghosts", namely states
with negative norm as one finds in QED when one quantizes the 
electromagnetic field in a covariant gauge. The consistency of the model 
requires the existence of relations satisfied by the scattering 
amplitudes that are similar to those obtained through gauge invariance 
in QED. If the model is consistent they must  decouple the negative 
norm states leaving us with a physical spectrum of positive norm
states. In order to 
study in a simple way these issues, we discuss in the next section 
the operator formalism introduced already in 
1969~\cite{FGV,NAMBU1,LENNY}.

Before  concluding this section let us go back to the non-planar 
four-point amplitude in Eq. (\ref{vir}) and  discuss its generalization 
to an $N$-point amplitude.  Using the technique of the electrostatic 
analogue  on the sphere instead of on the 
disk Shapiro~\cite{SHAPI}  was able to obtain a $N$-point amplitude
that reduces to the four-point amplitude in Eq. (\ref{vir}) with  intercept
$\alpha_0 =2$.  The $N$-point amplitude found in 
Ref.~\cite{SHAPI}  is:
\begin{eqnarray}
\int \frac{\prod_{i=1}^{N} d^2 z_i}{dV_{abc}} 
\prod_{i  <  j} | z_i - z_j |^{\alpha' p_i \cdot p_j}
\label{sv}
\end{eqnarray}
where
\begin{eqnarray}
d V_{abc} = \frac{d^2 z_a  d^2 z_b  d^2 z_c}{|z_a - z_b|^2 
|z_a - z_c |^2 | z_b - z_c |^2}
\label{me23}
\end{eqnarray}
The integral in Eq. (\ref{sv}) is performed in the entire complex plane.

\section{Operator formalism and factorization}
\label{ope}

The factorization properties of the dual resonance model were
first studied by factorizing by brute force the N-point amplitude
at the various poles~\cite{FUVE1,BAMA}.  The number of terms 
that factorize the residue of the pole at $\alpha (s) =n$, increases
rapidly with the value of $n$. In order to find their degeneracy it turned
out to be convenient  to first rewrite the N-point amplitude in
an operator formalism.
In this section we introduce the operator formalism and we rewrite the 
$N$-point amplitude derived in the previous section in this formalism.

The key idea~\cite{FGV,NAMBU1,LENNY} 
 is to introduce an infinite set of harmonic oscillators  
and a position 
and momentum operators~\footnote{Actually the position and 
momentum operators were introduced in Ref.~\cite{FUVE2}. } 
which satisfy the following commutation relations:
\begin{eqnarray}
[a_{n \mu} , a^{\dagger}_{m\nu} ] = \eta_{\mu \nu} \delta_{nm}~~~;~~~
[ {\hat{q}}_\mu , {\hat{p}}_\nu ] = i \eta_{\mu \nu} 
\label{osci}
\end{eqnarray}
where $\eta_{\mu \nu}$ is the flat Minkowski metric that we take to be
$ \eta_{\mu \nu} = (-1, 1, \dots 1)$.  A state with momentum $p$ is 
constructed in 
terms of a state with zero momentum as follows:
\begin{eqnarray}
{\hat{p}} | p \rangle \equiv  {\hat{p}}  {e}^{i p \cdot {\hat{q}}} | 
0 \rangle = p  | p \rangle ~~;~~
{\hat{p}}  \, |0\rangle  =0
\label{mom}
\end{eqnarray}
normalized as~\footnote{Although we now use an arbitrary $d$ we 
want to remind you that all original calculations  were done for $d=4$. }
\begin{eqnarray}
\langle p | p' \rangle = (2 \pi )^d \delta^{(d)} ( p+ p' )
\end{eqnarray}
In order to avoid minus signs we use the convention that
\begin{eqnarray}
\langle p | = \langle 0 | e^{i p \cdot {\hat{q}}}
\label{mom1}
\end{eqnarray}  
A complete and orthonormal basis of vectors in the harmonic oscillator space
 is given by
\begin{eqnarray}
| \lambda_1 , \lambda_2, \dots \lambda_i  ; p\rangle =  \prod_n
\frac{ ( a^{\dagger}_{\mu_n ;n})^{\lambda_{n; \mu_n}} 
}{\sqrt{\lambda_{n, \mu_n }!}}  
{e}^{i p {\hat{q}}} 
| 0, 0\rangle 
\label{comple}
\end{eqnarray}
where the first $| 0 \rangle$ corresponds to the one annihilated by
all annihilation operators and the second one to the state of zero momentum:
\begin{eqnarray}
a_{\mu_ n ;n } | 0 , 0 \rangle = {\hat{p}} | 0 , 0 \rangle =0
\end{eqnarray}
Notice that Lorentz invariance forces to introduce also oscillators that 
create states with negative norm due to the minus sign in the flat 
Minkowski metric. This implies that the space spanned by the 
states in Eq. (\ref{comple})  is not positive definite. This is, however, not 
allowed in a quantum theory and therefore if the dual resonance model 
is a consistent quantum-relavistic theory we expect the presence of relations
of the kind of those provided by gauge invariance in QED.

Let us  introduce the Fubini-Veneziano~\cite{FUVE2} operator:
\begin{eqnarray}
Q_{\mu} (z) = Q^{(+)}_{\mu} (z)  +  Q^{(0)}_{\mu}  (z) + Q^{(-)}_{\mu}  (z) 
\label{qfv}
\end{eqnarray}
where
\[
Q^{(+)} =  i \sqrt{2 \alpha'}  \sum_{n=1}^{\infty} 
\frac{a_n}{\sqrt{n}} z^{-n}~~;~~
Q^{(-)} =  -i \sqrt{2 \alpha'}  \sum_{n=1}^{\infty} 
\frac{a^{\dagger}_{n}}{\sqrt{n}}
 z^{ n}
 \]
 \begin{eqnarray}
 Q^{(0)} = {\hat{q}} - 2 i  \alpha' {\hat{p}} \log z
 \label{qfva} 
\end{eqnarray}
In terms of $Q$ we introduce the vertex operator corresponding to 
the external leg with momentum $p$:
\begin{eqnarray}
{{V}} ( z ; p ) = : {e}^{i p \cdot Q (z)} : \equiv   
{e}^{i p \cdot Q^{(-)} (z) }  
{e}^{i p {\hat{q}}}{e}^{ + 2 \alpha' {\hat{p}} 
\cdot p \log z}   {e}^{i p \cdot Q^{(+)} (z) }
\label{vertope}  
\end{eqnarray}
and compute the following vacuum expectation value:
\begin{eqnarray}
\langle 0, 0 | \prod_{i=1}^{N} {{V}} ( z_i , p_i)   |0, 0\rangle
\label{vev34}
\end{eqnarray}
It can be easily computed using the Baker-Haussdorf relation 
\begin{eqnarray}
e^A  e^B  =  e^B  e^A  e^{[A,B]}
\label{bh}
\end{eqnarray}
that is valid if the commutator, as in our case,  $[A,B]$ is a
c-number.  In our case
the commutation relations  to be used are:
\begin{eqnarray}
[ Q^{(+)} (z) , Q^{(-)} (w) ] = -2 \alpha' \log \left( 1- \frac{w}{z} \right)
\label{contra}
\end{eqnarray}
and  the second one in Eq. (\ref{osci}).  Using them one gets:
\begin{eqnarray}
{{V}} ( z; p ) {{V}} (w; k) = : {{V}} ( z; p ) {{V}} (w; k) :
(z-w)^{2 \alpha' p \cdot k}  
\end{eqnarray}
and
\begin{eqnarray}
\langle 0, 0 | \prod_{i=1}^{N} {{V}} ( z_i , p_i)   |0, 0\rangle =
 \prod_{i >j} ( z_i - z_j )^{2 \alpha' p_i \cdot p_j} (2 \pi)^d
 \delta^{(d)} ( \sum_{i=1}^{N} p_i )
\label{ANc}
\end{eqnarray}
where the normal ordering requires that all creation operators 
be put on the left of the annihilation one and the momentum operator
${\hat{p}}$ be put on the right of the position operator ${\hat{q}}$.
This means that
\[
( 2 \pi)^d \delta^{(d)}  ( \sum_{i=1}^{N}  p_i ) B_N = 
\int_{-\infty}^{\infty} \frac{\prod_{1}^{N} d z_i  
\theta (z_i - z_{i+1}) }{dV_{abc}}
\prod_{i=1}^{N} \left[ ( z_{i} - z_{i+1} )^{\alpha_0 -1} \right]
\times   
\]
\begin{eqnarray}
\times \langle 0, 0 | \prod_{i=1}^{N} {{V}} ( z_i , p_i)   |0, 0\rangle
\label{AN}
\end{eqnarray}
By choosing the three variables $z_a, z_b$ and $z_c$ as in
Eq. (\ref{cho8}) we can rewrite the previous equation as follows:
\[
( 2 \pi)^d \delta^{(d)}  ( \sum_{i=1}^{N}  p_i ) B_N = \int_{0}^{1}
\prod_{i=3}^{N-1}  d z_i    \prod_{i=2}^{N-1} \theta (z_{i}  - z_{i+1} )  
 \times 
\] 
\begin{eqnarray}
\times \prod_{i=2}^{N-1} \left[ ( z_{i} - z_{i+1} )^{\alpha_0 -1} \right]
\langle 0, p_1 | \prod_{i=2}^{N-1} V ( z_i ; p_i ) |
0, p_N \rangle
\label{bnf}
\end{eqnarray} 
where we have taken $z_2 =1$ and we have defined 
$( \alpha_0 \equiv \alpha' p_{i}^{2} ; i =1 \dots N)$ :
\begin{eqnarray}
\lim_{z_N \rightarrow 0} V (z_N ;p_N ) | 0,0\rangle \equiv | 0 ; p_N  \rangle
~~;~~ \langle 0; 0 | \lim_{z_1 \rightarrow \infty}  z_{1}^{2 \alpha_0}  
V (z_1 ; p_1 ) = \langle 0, p_1 |
\label{limi67}
\end{eqnarray} 
Before proceeding to factorize the $N$-point amplitude let us study 
the properties under the projective group of the operators 
that we have introduced.   We have already seen that
the projective group  leaves the integrand 
of the Koba-Nielsen representation of the $N$-point amplitude 
invariant.  The projective group  has three generators 
$L_0 , L_1$ and $L_{-1} $ corresponding respectively to 
dilatations, inversions and translations. 
Assuming that the Fubini-Veneziano fields $Q(z)$ 
transforms as a field with
weight 0  (as a scalar) we can immediately write the commutation 
relations that $Q (z)$ must satisfy.  This means in fact  
that, under a projective transformation, $Q(z)$
transforms as follows:
\begin{eqnarray}
Q (z) \rightarrow Q^T (z) = Q \left( \frac{\alpha z + 
\beta}{\gamma z + \delta} \right)~~~;~~~\alpha \delta - \beta \gamma =1
\label{qtra}
\end{eqnarray}
Expanding for small values of the parameters we get:
 \begin{eqnarray}
Q ^{T} (z) = Q(z) + ( \epsilon_1 + 
\epsilon_2  z + \epsilon_3  z^2  ) \frac{d Q (z)}{dz}
+ o (\epsilon^2 )
\label{traq}
\end{eqnarray}
This means that the three generators of the projective 
group must satisfy the following
commutation relations with $Q (z) $:
\begin{eqnarray}
[ L_0 , Q(z) ] = z \frac{dQ}{dz}~~;~~[L_{-1} , Q(z) ] = \frac{dQ}{dz}~~;~~
[L_{1} , Q(z) ] = z^2  \frac{dQ}{dz}
\label{commure}
\end{eqnarray}
They are given by the following expressions in terms of 
the harmonic oscillators:
\begin{eqnarray}
L_0 = \alpha' {\hat{p}}^2 + \sum_{n=1}^{\infty} n a_{n}^{\dagger} \cdot a_n~;~
L_1 = \sqrt{2 \alpha'} {\hat{p}} \cdot a_1 + 
\sum_{n=1}^{\infty} \sqrt{n(n+1)} a_{n+1} \cdot 
a_{n}^{\dagger}
\label{elle}
\end{eqnarray}
and 
\begin{eqnarray}
L_{-1} = L_{1}^{\dagger} = \sqrt{2 \alpha'} {\hat{p}} \cdot a_{1}^{\dagger} + 
\sum_{n=1}^{\infty} \sqrt{n(n+1)} a_{n+1}^{\dagger}  \cdot a_{n}
\label{L-1}
\end{eqnarray}
They annihilate the vacuum 
\begin{eqnarray}
L_0 | 0,0\rangle = L_1 |0,0 \rangle = L_{-1} | 0,0 \rangle =0
\label{vac}
\end{eqnarray}
that is therefore called the projective invariant vacuum, and  
satisfy the algebra
that is called Gliozzi 
algebra~\cite{GLIOZZI}\footnote{See also Ref.~\cite{CMR}.}:
\begin{eqnarray}
[L_0 , L_1] = - L_1~~;~~[L_0 , L_{-1} ] = L_{-1}~~;~~[L_1 , L_{-1} ] = 2 L_0
\label{nando}
\end{eqnarray}
The vertex operator with momentum $p$ is a projective field with weight equal to 
$\alpha_0 = \alpha' p^2$.  It transforms 
in fact as follows under the projective group:
\begin{eqnarray}
[L_n , V (z, p) ] = z^{n+1} \frac{d V (z, p)}{dz} + \alpha_0 
(n +1) z^n V (z, p )~~;~~n=0, \pm 1
\label{prov}
\end{eqnarray}
or in finite form as follows:
\begin{eqnarray}
U V ( z , p ) U^{-1} = \frac{1}{(\gamma z + \delta)^{2\alpha_0 }} 
V \left( \frac{\alpha z + \beta}{\gamma z + \delta } , p \right)
\label{arb}
\end{eqnarray}
where $U$ is the generator of an arbitrary finite projective transformation.   

Since $U$ leaves the vacuum invariant,  by 
using Eq. (\ref{arb}) it is easy to show that:
\begin{eqnarray}
\langle 0,0 |  \prod_{i=1}^{N} V (z_{i} ' , p ) | 0,0 \rangle = 
\prod_{i=1}^{N} {(\gamma  z_{i} + \delta)^{2\alpha_0 }} \langle 0, 0| 
\prod_{i=1}^{N} V (z_{i} , p ) |0,0 \rangle     
\label{tra45}
\end{eqnarray}
that together with the following equation:
\begin{eqnarray}
\prod_{i=1}^{N} d z_i ' \prod_{i=1}^{N} ( z_{i}' - z_{i+1} ')^{\alpha_0  -1} = 
\prod_{i=1}^{N} d z_i  \prod_{i=1}^{N-1} ( z_{i} - z_{i+1} )^{\alpha_0  -1} 
\prod_{i=1}^{N} ( \gamma z_{i} + \delta )^{-2 \alpha_0}
\label{tra46}
\end{eqnarray}
implies that the integrand of the $N$-point amplitude in Eq. (\ref{AN}) is 
invariant under projective transformations.

We are now ready to factorize the $N$-point amplitude and find the 
spectrum of mesons. 

{From} Eq.s (\ref{prov}) and (\ref{arb}) it is easy to derive the
transformation of the vertex operator under a finite dilatation:
\begin{eqnarray}
z^{L_0} V (1  , p) z^{-L_0} = V ( z, p) z^{\alpha_0}
\label{dila4}
\end{eqnarray}
Changing the integration variables as follows:
\begin{eqnarray}
x_i = \frac{z_{i+1}}{z_i}~~;~~i=2, 3 \dots N-2~~~;~~~ \det \frac{\partial
  z_i}{\partial x_j} = z_3 z_4 \dots z_{N-2}
\label{newva6}
\end{eqnarray}
where the last term is the jacobian of the trasformation from $z_i$ to
$x_i$, we get from Eq.(\ref{bnf}) the following expression:
\begin{eqnarray}
A_{N} \equiv 
\langle 0, p_1 | V (1, p_2) D V ( 1, p_3 ) \dots  D V (1, p_{N-1} ) |
0, p_N \rangle
\label{bng}
\end{eqnarray}
 where the propagator $D$ is equal to:
\begin{eqnarray}
D = \int_{0}^{1} d x x^{L_0 -1 - \alpha_0 } (1 -x )^{\alpha_0 -1} = 
\frac{  \Gamma ( L_0 - \alpha_{0} ) \Gamma ( \alpha_{0}  )}{
\Gamma ( L_0  ) }
\label{propa5}
\end{eqnarray}
and
\begin{eqnarray}
A_{N} = (2 \pi )^d \delta^{(d)} \left( \sum_{i=1}^{N} p_i \right)B_N
\label{AB}
\end{eqnarray}
The factorization properties of the amplitude can be studied by
inserting in the channel $(1, M)$ or equivalently in the channel $(M+1
, N)$ described by the Mandelstam variable 
\begin{eqnarray}
s = - (p_1 + p_2 + \dots p_M)^2  = - (p_{M+1} + p_{M+2} \dots + p_{N})^2
\equiv - P^2
\label{es56}
\end{eqnarray}
the complete set of states given in Eq. (\ref{comple}):
\begin{eqnarray}
A_N =  \sum_{\lambda, \mu} \langle p_{(1, M)}  | \lambda, P \rangle
\langle \lambda, P | D | \mu, P \rangle \langle \mu, P | p_{(M+1, N)} 
 \rangle 
\label{facta}
\end{eqnarray}
where
\begin{eqnarray}
\langle p_{(1, M)}   | = 
\langle 0, p_1 | V (1, p_2) D V ( 1, p_3 ) \dots V (1, p_M )
\label{factc}
\end{eqnarray}
and
\begin{eqnarray}
| p_{(M+1, N) }   \rangle =  
V (1, p_{M+1} ) D \dots V ( 1, p_{N-1} ) | p_{N} ,0\rangle  
\label{factb}
\end{eqnarray}
Introducing the quantity:
\begin{eqnarray}
R = \sum_{n=1}^{\infty} n a_{n}^{\dagger} \cdot a_n
\label{erre}
\end{eqnarray}
it is possible to rewrite
\begin{eqnarray}
\langle \lambda, P | D | \mu, P \rangle = \sum_{m=0}^{\infty}
\langle \lambda, P |
\frac{(-1)^m \left(\begin{array}{c} \alpha_0 -1 \\
                                     m \end{array}  \right)}{R + m -
                                 \alpha (s)}| \mu, P \rangle
\label{factd}
\end{eqnarray}
where $s$ is the variable defined in Eq. (\ref{es56}). Using this
equation we can rewrite Eq. (\ref{facta}) as follows
\begin{eqnarray}
A_N =  \sum_{\lambda, \mu  } \langle p_{(1, M)}   | \lambda, P
\rangle 
\sum_{m=0}^{\infty}
\langle \lambda, P |
\frac{(-1)^m \left(\begin{array}{c} \alpha_0 -1 \\
                                     m \end{array}  \right)}{R + m -
                                 \alpha (s)}| \mu, P \rangle
\langle \mu, P | p_{(M+1, N)}   \rangle 
\label{factf}
\end{eqnarray}
This expression
shows that amplitude $A_N$ has a pole in the channel $(1, M)$ when 
$\alpha (s)$  is equal to an integer $n \geq 0$ and the states $| \lambda
\rangle$ that contribute to its residue are those satisfying the relation: 
\begin{eqnarray}
R | \lambda \rangle  = (n-m) | \lambda \rangle ~~~;~~~m=0, 1  \dots n
\label{pole}
\end{eqnarray}
The number of independent states $| \lambda \rangle$ contributing to
the residue gives the degeneracy of states for each level $n$.

Because of manifest relativistic invariance the space spanned by the
complete system of states in Eq. (\ref{comple}) contains states with 
negative norm corresponding to those states having an odd number of
oscillators with timelike directions (see Eq. (\ref{osci})). This is
not consistent in a quantum theory where the states of a system must
span a positive definite Hilbert space. This means that there must
exist a number of relations satisfied by the external states that
decouple a number of states leaving with a positive definite Hilbert
space. In order to find these relations we rewrite the state 
in Eq. (\ref{factb}) going back to the Koba-Nielsen variables:
\[
| p_{(1, M)}  \rangle = \prod_{i=2}^{M-1} [\int  d z_i \theta (
  z_i - z_{i+1} )] \prod_{i=1}^{M-1} ( z_i - z_{i+1} )^{\alpha_0 -1}
  \times  
\]
\begin{eqnarray}
\times 
V (1, p_1) V ( z_2 , p_2 ) \dots V ( z_{M-1} , p_{M-1} ) | 0, p_{M} \rangle 
\label{pstate}
\end{eqnarray}
Let us  consider the operator  $ U (\alpha )$ that generate the  projective 
transformation that leaves the points $z=0, 1$ invariant:
\begin{eqnarray}
z' = \frac{z}{1 - \alpha (z-1)} = z + \alpha ( z^2 - z ) + o( \alpha^2 )
\label{prota}
\end{eqnarray}
From the transformation properties of the vertex operators in
Eq. (\ref{arb}) it is easy to see that the previous transformation
leaves the state in Eq. (\ref{pstate}) invariant:
\begin{eqnarray}
 U (\alpha ) | p_{(1, M)}   \rangle = | p_{(1, M)}   \rangle
\label{inva41}
\end{eqnarray}
This means that the generator of the previous transformation
annihilates the state in Eq. (\ref{pstate}):
\begin{eqnarray}
W_1 | p_{(1, M)}   \rangle =0~~~;~~W_1 = L_1 - L_0
\label{gene}
\end{eqnarray}
The explicit form of $W_1$ follows from the infinitesimal form of the
transformation in Eq. (\ref{prota}). This condition that is of the
same kind of the relations that on shell amplitudes with the emission
of photons satisfy as a consequence of gauge invariance, implies that the
residue at the pole in Eq. (\ref{factf}) can be factorized with a smaller
number of states. It turns out, however, that a detailed analysis of
the spectrum shows that negative norm states are still present. This
can be qualitatively understood as follows. Due to the Lorentz metric 
we have a negative norm component for each oscillator. In order to be
able to decouple all negative norm states we need to have a gauge
condition of the type as in Eq. (\ref{gene}) for each oscillator. But
the number of oscillators is infinite and, therefore, we need an
infinite number of conditions of the type as in Eq. (\ref{gene}). It
was found in Ref.~\cite{VIRA2} that, if we take $\alpha_0 =1$, then one can
easily construct an infinite number of operators that leave the state
in Eq. (\ref{pstate}) invariant. In the next section  we will 
concentrate on this case. 

\section{The case $\alpha_0 =1$}
\label{sec4}

If we take $\alpha_0 =1$ many of the formulae given in the previous 
section simplify. The $N$-point amplitude in Eq. (\ref{bnd}) becomes: 
\begin{eqnarray}
B_N = \int_{-\infty}^{\infty} \frac{\prod_{1}^{N} d z_i  
\theta (z_i - z_{i+1}) }{dV_{abc}}
\prod_{j>i} ( z_i -
z_j )^{2 \alpha' p_i \cdot p_j}
\label{bnd1}
\end{eqnarray}
that can be rewritten in the operator formalism as follows:
\begin{eqnarray}
( 2 \pi)^4 \delta ( \sum_{i=1}^{N}  p_i ) B_N = 
\int_{-\infty}^{\infty} \frac{\prod_{1}^{N} d z_i  
\theta (z_i - z_{i+1}) }{dV_{abc}}
 \langle 0, 0 | \prod_{i=1}^{N} {{V}} ( z_i , p_i)   |0, 0\rangle
\label{AN1}
\end{eqnarray}
By choosing $z_1 = \infty, z_2 =1$ and $z_N =0$ it becomes
\[
( 2 \pi)^4 \delta ( \sum_{i=1}^{N}  p_i ) B_N = 
\]
\begin{eqnarray}
=\int_{0}^{1}
\prod_{i=3}^{N-1}  d z_i    \prod_{i=2}^{N-1} \theta (z_{i}  - z_{i+1} )  
\langle 0, p_1 | \prod_{i=2}^{N-1} V ( z_i ; p_i ) |
0, p_N \rangle
\label{bnfb}
\end{eqnarray} 
where
\begin{eqnarray}
\lim_{z_N \rightarrow 0} V (z_N ;p_N ) | 0,0\rangle \equiv | 0 ; p_N \rangle
~~;~~ \langle 0; 0 | \lim_{z_1 \rightarrow \infty}  z_{1}^{2 }  
V (z_1 ; p_1 ) = \langle 0, p_1 |
\label{limi671}
\end{eqnarray} 
Eq. (\ref{bng}) is as before, but now the propagator becomes:
\begin{eqnarray}
D = \int dx x^{L_0 -2} = \frac{1}{L_0 -1}
\label{proab}
\end{eqnarray}
This means that Eq. (\ref{factd}) becomes:
\begin{eqnarray}
\langle \lambda, P | D | \mu, P\rangle = \langle \lambda, P | \frac{1}{L_0 -1} 
| \mu, P \rangle
\label{pro82}
\end{eqnarray}
and Eq. (\ref{factf}) has the simpler form:
\begin{eqnarray}
B_N =  \sum_{\lambda } \langle p_{(1, M)}  | \lambda, P \rangle 
\langle \lambda, P | \frac{1}{R -  \alpha (s)}| \lambda, P \rangle
\langle \lambda, P | p_{(M+1, N)}   \rangle  
\label{factf1}
\end{eqnarray}
$B_N$ has a pole in the channel $(1, M)$ when $\alpha (s)$ is equal to 
an integer $n \geq 0$ and the states $| \lambda \rangle$ that 
contribute to its residue are those satisfying the relation:
\begin{eqnarray}
R | \lambda \rangle = n | \lambda \rangle
\label{dege}
\end{eqnarray}
Their number gives the degeneracy of the states contributing to 
the pole at $\alpha (s) =n$. 
The  $N$-point amplitude can be written as:
\begin{eqnarray}
B_N = \langle p_{(1, M )}   | D | p_{(M+1, N)}   \rangle
\label{anbi}
\end{eqnarray}
where
\[
| p_{(1,M)}   \rangle = \int \prod_{i=2}^{M-1} 
\left[ dz_i \theta (z_i - z_{i+1} )\right] \times
\]
\begin{eqnarray}
\times
V( 1, p_1 ) V (z_2 , p_2 ) \dots V ( z_{M-1} , p_{M-1} | 0 , p_{M} \rangle
\label{psta4}
\end{eqnarray}
Using Eq. (\ref{dila4})  and changing variables from $z_i , i=2 \dots M-1$ 
to $x_i = \frac{z_{i+1} }{z_i  }, 
i=1 \dots M-2 $ with $z_1=1$ we can rewrite the previous equation as follows:
\begin{eqnarray}
| p_{(1,M)}   \rangle = V( 1, p_1 ) 
D V( 1, p_2 ) \dots D V( 1, p_{M-1} )  | 0 , p_{M} \rangle
\label{psta5}
\end{eqnarray}
where the propagator $D$ is defined in Eq. (\ref{proab}).

We want now to show that the state in 
Eq.s (\ref{psta4}) and (\ref{psta5}) is not only 
annihilated by the operator in 
Eq. (\ref{gene}), but, if $\alpha_0 =1$~\cite{VIRA2},  
by an infinite set of operators
whose lowest one is the one in Eq. (\ref{gene}).  We will derive this by 
using the formalism developed in Ref.~\cite{FUVE3} and we will
follow closely their derivation.

Starting from Eq.s (\ref{commure}) Fubini and Veneziano 
realized that the generators 
of the projective group acting  on  a function of z are given by:
\begin{eqnarray}
L_0 = - z \frac{d}{dz}~~;~~L_{-1} =  -\frac{d}{dz}~~;~~L_1 = -z^2 \frac{d}{dz}
\label{genpro}
\end{eqnarray}
They generalized the previous generators to an arbitrary conformal 
transformation by introducing the following operators, called 
Virasoro operators:
\begin{eqnarray}
L_n = -z^{n+1} \frac{d}{dz}
\label{confge}
\end{eqnarray}
that satisfy the algebra:
\begin{eqnarray}
[ L_n , L_m ] = (n-m) L_{n+m}
\label{viralg}
\end{eqnarray}
that does not contain the term with the central charge!
They also showed that  the Virasoro operators 
 satisfy the following commutation 
relations with the vertex operator:
\begin{eqnarray}
[ L_n , V (z , p) ]  = \frac{d}{dz} \left( z^{n+1} V (z , p) \right)
\label{commu6}
\end{eqnarray}
More in general actually they  define an operator  $L_f$ corresponding to an 
arbitrary function $f ( \xi )$ and $L_f  = L_n$ if we 
choose $f(\xi ) = \xi^{n} $. In this case the commutation relation in
Eq. (\ref{commu6}) becomes:
\begin{eqnarray}
[ L_f , V (z , p) ]  = \frac{d}{dz} \left( z f(z) V (z , p) \right)
\label{commu6c}
\end{eqnarray}
By introducing the variable:
\begin{eqnarray}
y = \int_{A}^{z} \frac{d \xi}{\xi f( \xi )}
\label{newva4}
\end{eqnarray}
where $A$ is an arbitrary constant, 
one can rewrite Eq. (\ref{commu6c}) in the following form:
\begin{eqnarray}
[ L_f , z f (z) V (z , p) ]  = \frac{d}{dy} \left( z f(z) V (z , p) \right)
\label{commu6d}
\end{eqnarray}
This implies that, under an arbitrary conformal 
transformation  $z \rightarrow f(z)$, generated by $U = {e}^{\alpha L_f}$, 
the vertex operator transforms as:
\begin{eqnarray}
{e}^{\alpha L_f } V ( z, p )\,   z f(z)\, {e}^{- \alpha L_f} =  
V ( z' ,p ) z' f( z' )
\label{confoa} 
\end{eqnarray}
where the parameter $\alpha$ is given by:
\begin{eqnarray}
\alpha = \int_{z}^{z'} \frac{d \xi}{\xi f (\xi)}
\label{alpha}
\end{eqnarray}
On the other hand, this equation implies:
\begin{eqnarray}
\frac{dz}{z f (z) } = \frac{dz' }{z' f (z') }
\label{dz}
\end{eqnarray}
that, inserted in Eq. (\ref{confoa}), implies that 
the quantity $V (z, p) \, dz$ is left invariant by the 
transformation $z \rightarrow f(z)$:
\begin{eqnarray}
{e}^{\alpha L_f } V ( z, p )  dz {e}^{- \alpha L_f} =  V ( z' ,p ) d z'
\label{confob}
\end{eqnarray}
Let us now act with the previous conformal transformation 
on the state in Eq. (\ref{psta4}). We get:
\[
{e}^{\alpha L_f }  | p_{(1, M)} \rangle =  \int_{0}^{1}  \prod_{i=2}^{M-1} 
\left[ d z_i \theta (z_i - z_{i+1} ) \right]
{e}^{\alpha L_f }  V (1 , p_1 ) {e}^{-\alpha L_f }  \times
\]
\[
\times
{e}^{\alpha L_f }  V (z_2 , p_2 )  {e}^{- \alpha L_f } \dots 
\dots {e}^{\alpha L_f }  V (z_{M-1}  , p_{M-1} )  {e}^{- \alpha L_f }
{e}^{\alpha L_f } | 0, p_{M} \rangle = 
\]
\[
= \int_{0}^{1} \prod_{i=2}^{M-1} \theta (z_i - z_{i+1}) \times 
{e}^{\alpha L_f }  V (1 , p_1 ) {e}^{-\alpha L_f } \times
\]
\begin{eqnarray}
\times
V ( z_{2}' , p_2 ) dz_{2}' \dots  
V ( z_{M-1}' , p_{M-1} ) dz_{M-1}'    {e}^{\alpha L_f }  | 0, p_{M} \rangle
\label{act45}
\end{eqnarray}
where we have used Eq. (\ref{confob}).  The previous 
transformation leaves the state invariant if both $z=0$ and $z=1$ are 
fixed points of 
the conformal transformation. This happens if the denominator 
in Eq. (\ref{alpha})   vanishes when $\xi =0, 1$. This requires the 
following conditions:
\begin{eqnarray}
f (1) =0~~~~;~~~\lim_{\xi \rightarrow 0} \xi f ( \xi ) =0
\label{fix}
\end{eqnarray}
Expanding $\xi$  near the poinr $\xi =1$ we can determine the relation between
$z$ and $z'$ near $z=z' =1$. We get:
\begin{eqnarray}
z' = \frac{z {e}^{- \alpha  f' (1)}}{1 - z + z {e}^{-\alpha f' (1)}}
\label{xi=1} 
\end{eqnarray}
and from it we can determine the conformal factor:
\begin{eqnarray}
\frac{dz'}{dz} = \frac{{e}^{- \alpha  
f' (1)}}{( 1 -z + z {e}^{- \alpha  f' (1)} )^2}
 \rightarrow   {e}^{ \alpha  f' (1)}
 \label{xi=1b}
\end{eqnarray}
in the limit $z \rightarrow 1$.
Proceeding in the same near the point $z=z' =0$ we get:
\begin{eqnarray}
z' = \frac{ z f(0) {e}^{\alpha f(0) }}{ 
f(0) + z f' (0) ( 1- {e}^{\alpha f(0) }} \rightarrow
z {e}^{\alpha f(0) }
\label{xi=0} 
\end{eqnarray}
in the limit $ z \rightarrow 0$.  This means that Eq. (\ref{act45}) becomes
\begin{eqnarray}
{e}^{\alpha \left(  L_f  - f' (1) - f(0) \right) }  | p_{(1, M)} 
\rangle =  | p_{(1 , M)} \rangle
\label{ident}
\end{eqnarray}
A choice of $f$ that satisfies Eq.s (\ref{fix}) is the following:
\begin{eqnarray}
f ( \xi ) =  \xi^n -1 
\label{cho8a}
\end{eqnarray}
that gives the following gauge operator:
\begin{eqnarray}
W_n = L_n - L_0 - (n-1)
\label{wn} 
\end{eqnarray}
that annihilates the state in Eq. (\ref{psta4}):
\begin{eqnarray}
W_n  |p_{1 \dots M}  \rangle = 0~~;~~n=1 \dots \infty
\label{gau78}
\end{eqnarray}
These are the  Virasoro conditions found in Ref.~\cite{VIRA2}. There is one 
condition for each negative norm oscillator and, therefore, in this case there
is the possibility that the physical subspace is positive definite. 
An alternative 
more direct derivation of Eq. (\ref{gau78}) can be obtained by acting 
with $W_n$ on the state in Eq. (\ref{psta5}) 
and using the following identities:
\begin{eqnarray}
W_n V (1, p) = V(1,p) ( W_n  +n )~~;~~( W_n  +n )D = [L_0 +n -1]^{-1} W_n
\label{sider1}
\end{eqnarray}
The second equation is a consequence of the following equation:
\begin{eqnarray}
L_n x^{L_0} = x^{L_0 +n} L_n
\label{rel83}
\end{eqnarray}
Eq.s (\ref{sider1}) imply 
\begin{eqnarray}
W_n  V (1, p) D = V(1,p) [L_0  +n -1]^{-1}  W_n
\label{sider2b}
\end{eqnarray}
This shows that the operator $W_n$ goes unchanged through all the product
of terms $VD$ until it  arrives in  front of the term 
$ V (1, p_{M-1}) | 0, p_M \rangle$.
Going through the vertex operator it becomes $L_n -L_0 +1$ that 
then annihilate  
the state
\begin{eqnarray}
(L_n -  L_0  +1 ) | p_M , 0  \rangle =0
\label{sider2}
\end{eqnarray}
This proves Eq. (\ref{gau78}).
  
Using the representation of the Virasoro operators given in Eq. (\ref{confge}) 
Fubini and Veneziano showed that they 
satisfy the algebra given in eq. (\ref{viralg})
without the central charge.  The presence of the  central charge 
was recognized by Joe 
Weis\footnote{See noted added in proof in Ref.~\cite{FUVE3}.}
 in 1970 and never published. Unlike Fubini and Veneziano~\cite{FUVE3} he used
 the expression of the $L_n$ operators in terms of the harmonic oscillators:  
\[
L_n = \sqrt{2 \alpha' n} {\hat{p}} \cdot a_{n} + \sum_{m=1}^{\infty} 
\sqrt{m (n+m)} a_{n+m} \cdot a_{m}  +
\]
\begin{eqnarray}
+ \frac{1}{2} 
\sum_{m=1}^{n-1} \sqrt{m ( n-m)} a_{m-n} \cdot a_m~~~;
n \geq 0~~~L_n = L_{n}^{\dagger} 
\label{ellen}
\end{eqnarray}
He got the following algebra:
\begin{eqnarray}
[L_n , L_m ] = (n-m) L_{n+m} +  \frac{d}{24} n  (n^{2} -1 ) \delta_{n+m;0}
\label{lnlm}
\end{eqnarray} 
where $d$ is the dimension of the Minkowski space-time. We write here $d$
for the dimension of the Minkowski space, but we want to remind you
that almost everybody working in a model for mesons at that time  took for 
granted that the dimension of the space-time was $d=4$. 
As far as I  remember the first
paper where a dimension $d\neq 4$ was introduced was Ref.~\cite{LOVE1} where 
it was shown that the unitarity violating cuts in the non-planar 
loop become poles that were consistent with unitarity if $d=26$.     

In the last part of this section we will generalize the factorization 
procedure to the Shapiro-Virasoro model whose $N$-point amplitude 
is given in Eq. (\ref{sv}). In this case  we must introduce two sets 
of harmonic oscillators commuting with each other and only one set 
of zero modes satisfying the algebra~\cite{DELGIU2} :
\begin{eqnarray}
[a_{n \mu} , a^{\dagger}_{m\nu} ] =  
[{\tilde{a}}_{n \mu} , {\tilde{a}}^{\dagger}_{m\nu} ] = 
\eta_{\mu \nu} \delta_{nm}~~;~~ 
[ {\hat{q}}_\mu , {\hat{p}}_\nu ] = i \eta_{\mu \nu} 
\label{osci52}
\end{eqnarray}
In terms of them we can introduce the Fubini-Veneziano 
operator
\[
Q (z, {\bar{z}} )= {\hat{q}} - 2 \alpha' {\hat{p}} \log (z {\bar{z}}) +
i \frac{ \sqrt{2 \alpha'}}{2} \sum_{n=1}^{\infty} \frac{1}{\sqrt{n}} 
\left[ a_{n} z^{-n} - a_{n}^{\dagger} z^n \right] +
\]
\begin{eqnarray}
+ i \frac{ \sqrt{2 \alpha'}}{2} \sum_{n=1}^{\infty} \frac{1}{\sqrt{n}} 
\left[ {\tilde{a}}_{n} {\bar{z}}^{-n} - 
{\tilde{a}}_{n}^{\dagger} {\bar{z}}^n \right] 
\label{QFV}
\end{eqnarray}
We can then introduce the vertex operator:
\begin{eqnarray}
{{V}} ( z, {\bar{z}} ; p ) = : {e}^{i p \cdot Q (z, {\bar{z}})} : 
\label{veope}
\end{eqnarray}
and write the $N$-point amplitude in Eq. (\ref{gene}) 
in the following factorized form:
\[
\int \frac{\prod_{i=1}^{N} d^2 z_i}{dV_{abc}}  
\langle 0 | R \left[ \prod_{i=1}^{N} 
V (z_i , {\bar{z}}_i , p_i ) ) \right] |0 \rangle  = 
\]
\begin{eqnarray}
= (2 \pi)^4 \delta^{(4)} ( \sum_{i=1}^{N} p_i )
\int \frac{\prod_{i=1}^{N} d^2 z_i}{dV_{abc}} 
\prod_{i  <  j} | z_i - z_j |^{\alpha' p_i \cdot p_j}
\label{fact}
\end{eqnarray}
where the radial ordered product is given by
\begin{eqnarray}
R \left[ \prod_{i=1}^{N}  V (z_i , {\bar{z}}_i , p_i ) ) \right] =
\prod_{i=1}^{N}  V (z_i , {\bar{z}}_i , p_i ) ) 
\prod_{i=1}^{N-1} \theta (|z_i  | - |z_{i+1} | )
+ \dots  
\label{radi}
\end{eqnarray}
and the dots indicate a sum over all permutations 
of the vertex operators.

By fixing $z_1 = \infty, z_2 =1 , z_N =0$ we can 
rewrite the previous expression as follows:
\begin{eqnarray}
\int \prod_{i=3}^{N-1} d^2 z_i   
\langle 0, p_1 | R \left[ \prod_{i=2}^{N-1} 
V (z_i , {\bar{z}}_i , p_i ) ) \right] |0, p_N \rangle   
\label{fact4}
\end{eqnarray}
For the sake of simplicity let us consider the term corresponding to the
permutation $1,2, \dots N$. In this case the Koba-Nielsen variables are 
ordered in such a way that $|z_i| \geq |z_{i+1}|$ for $i=1,\dots N-1$. We can 
then use the formula:
\begin{eqnarray}
V (z_i , {\bar{z}}_i , p_i ) ) = 
z_{i}^{L_0 -1} {\bar{z}_i}^{{\tilde{L}}_0  -1} V(1,1, p_i ) 
z_{i}^{-L_0} {\bar{z}_i}^{-{\tilde{L}}_0} 
\label{dila56}
\end{eqnarray}
and change variables:
\begin{eqnarray}
w_i = \frac{ z_{i+1}}{z_i}~~;~~|w_i  | \leq 1
\label{ww}
\end{eqnarray}
to rewrite   Eq. (\ref{fact4}) as follows:
\begin{eqnarray}
\langle 0, p_1 | V(1,1, p_i1) D 
V(1,1, p_2 ) D \dots   V(1,1, p_{N-1} ) |0, p_N \rangle  
\label{fac52}
\end{eqnarray}
where
\begin{eqnarray}
D = \int \frac{ d^2 w}{|w|^2} \,\, w^{L_0  -1} {\bar{w}}^{{\tilde{L}}_0 -1} = 
\frac{2}{L_0 + {\tilde{L}}_0 -2 } \cdot
\frac{\sin \pi (L_0 - {\tilde{L}}_0 )}{L_0 - {\tilde{L}}_0} 
\label{propa45}
\end{eqnarray}
We can now follow the same procedure for all permutations  arriving at
the following expression:
\begin{eqnarray}
\langle 0, p_1 | P [ V(1,1, p_2 ) D V(1,1, p_3 ) D \dots   V(1,1, p_{N-1} ) ] 
|0, p_N \rangle  
\label{permu}
\end{eqnarray}
where P means a sum of all permutations of the particles.

If we want to consider the factorization of the amplitude on the pole at 
$s =- (p_1 + \dots p_M )^2$ we get only the following contribution:
\begin{eqnarray}
 \langle  p_{(1 \dots M)} |D | p_{(M+1 \dots N)} \rangle
\label{fac52b}
\end{eqnarray}
where
\begin{eqnarray}
 | p_{(M+1 \dots N)} \rangle = P [
V (1,1, p_{M+1} ) D \dots V (1,1, p_{N-1} ] |0, p_N \rangle 
\label{p89}
\end{eqnarray}
and
\begin{eqnarray}
 \langle  p_{(1 \dots M)} | = \langle 0, p_1 | 
P \left[ V (1,1, p_2 ) D \dots V (1,1, p_M)
\right] 
\label{p88}
\end{eqnarray}
The amplitude is factorized by introducing a complete set of 
states and rewriting
Eq. (\ref{fac52}) as follows:
\begin{eqnarray}
\sum_{\lambda, {\tilde{\lambda}}}  \langle p_{1 \dots M} | 
\lambda , {\tilde{\lambda}} \rangle     \frac{ 2 \pi 
\langle \lambda , {\tilde{\lambda} } | \delta_{L_0 , {\tilde{L}}_0}
|\lambda , {\tilde{\lambda}} \rangle    }  { L_0 + {\tilde{L}}_0 - 2} 
\langle \lambda , {\tilde{\lambda}} |  p_{(M+1, \dots N)} \rangle
\label{fac53}
\end{eqnarray}
By writing
\begin{eqnarray}
L_0 = \frac{\alpha'}{4} {\hat{p}}^2 + R ~~;~~ {\tilde{L}}_0 = 
\frac{\alpha'}{4} {\hat{p}}^2 + {\tilde{R}} 
\label{L0Ltil0}
\end{eqnarray}
with
\begin{eqnarray}
R = \sum_{n=1}^{\infty} n a_{n}^{\dagger} \cdot a_n~~;~~
{\tilde{R}} = \sum_{n=1}^{\infty} n {\tilde{a}}_{n}^{\dagger} 
\cdot {\tilde{a}}_n
\label{NtildeN}
\end{eqnarray}
we can rewrite Eq. (\ref{fac53}) as follows
\begin{eqnarray}
\sum_{\lambda, {\tilde{\lambda}}}  \langle p_{1 
\dots M} | \lambda , {\tilde{\lambda}} \rangle     
\frac{ 2 \pi \langle \lambda , {\tilde{\lambda} } | \delta_{R , {\tilde{R}}}
|\lambda , {\tilde{\lambda}} \rangle    }  { R + {\tilde{R}} -  \alpha (s)} 
\langle \lambda , {\tilde{\lambda}} |  p_{(M+1, \dots N)} \rangle
\label{fact38}
\end{eqnarray}
We see that  the amplitude for the Shapiro-Virasoro model has 
simple poles only for even integer values of  $ 
\alpha_{SV} (s) = 2 + \frac{\alpha'}{2}  s = 2n  \geq 0 $ 
and the residue at the poles
factorizes in a sum with a finite number of terms.
Notice that the Regge trajectory of the Shapiro-Virasoro model 
has double intercept and half slope of that of the 
generalized Veneziano model. 

\section{Physical states and their vertex operators}
\label{phys}

In the previous section, we have seen that the residue at the 
poles of the $N$-point amplitudes factorizes in a sum of a 
finite number of terms. We have also seen 
that some of these terms, due to the Lorentz metric,  
correspond to states with negative norm. We have also derived
a number of  "Ward identities" given in Eq. (\ref{gau78}) 
that imply that some of the terms of the residue decouple.  
The question to be answered now is: 
Is the space spanned by the physical states a 
positive norm Hilbert space? In order to answer  
this question we need  first to find the conditions that 
characterize the on shell physical states 
$|\lambda, P \rangle$ and then to determine 
which are the states that   
contribute to the residue of the pole at 
$\alpha (s = - P^2) =n$.  In other words, we
have to find a way of characterizing the physical states
 and  of eliminating the spurious states that
decouple in Eq. (\ref{factf1}) as a consequence of Eq.s (\ref{gau78}).
A state $|\lambda. P\rangle $ contributes at the residue of the pole
in Eq.(\ref{factf1}) for $\alpha (s = - P^2) =n$ 
if it is on shell, namely if it satisfies the
following equations:
\begin{eqnarray}
R | \lambda , P \rangle = n | \lambda , P \rangle~~;~~ \alpha ( -P^2 )
= 1 - \alpha' P^2= n
\label{onshe}
\end{eqnarray} 
that can be written in a unique equation:
\begin{eqnarray}
(L_0 -1 ) | \lambda , P \rangle =0
\label{L0-1}
\end{eqnarray} 
Because of Eq. (\ref{gau78}) we also know that a state of the type:
\begin{eqnarray}
| s, P \rangle = W_{m}^{\dagger} | \mu , P \rangle
\label{spu}
\end{eqnarray} 
is not going to contribute to the residue of the pole. We call it a
spurious or unphysical state.  
We start constructing   the subspace of spurious states 
that are on shell at the level $n$.  Let us consider the set of 
orthogonal states
$ | \mu , P \rangle$ such that
\begin{eqnarray}
R | \mu , P \rangle = n_{\mu} | \mu , P \rangle~~;~~
 L_0 | \mu , P \rangle = (1-m) | \mu , P \rangle~~;~~1 - \alpha' P^2 =n
\label{onshe67}
\end{eqnarray} 
where
\begin{eqnarray}
m = n+ n_{\mu}
\label{mn}
\end{eqnarray} 
In terms of these states we can construct the most general spurious
state that is on shell at the level $n$. It is given by
\begin{eqnarray}
| s, P \rangle = W_{m}^{\dagger} | \mu , P \rangle~~;~~
(L_0 - 1) | s , P \rangle = 0
\label{spu4}
\end{eqnarray} 
per any positive integer $m$.  Using Eq. (\ref{onshe67}), 
eq. (\ref{spu4}) becomes:
\begin{eqnarray}
 | s , P \rangle = L_{m}^{\dagger} | \mu , P \rangle
\label{spu83}
\end{eqnarray} 
where $ | \mu , P \rangle$ is an arbitrary state satisfying 
Eq.s (\ref{onshe67}).

A physical state $|\lambda , P \rangle$ is defined as the one that is
orthogonal to all spurious states appearing at a certain level
$n$. This means that it must satisfy the following equation:
\begin{eqnarray}
\langle \lambda . P |  L_{\ell}^{\dagger} | \mu , P \rangle =0 
\label{phys9}
\end{eqnarray} 
for any state $| \mu , P \rangle$ satisfying Eq.s
(\ref{onshe67}). In conclusion,  the on shell physical states at the
level $n$ are characterized by the fact that
they  satisfy the following conditions:
\begin{eqnarray}
L_{m} | \lambda, P \rangle = (L_0 -1) | \lambda, P \rangle
=0~~;~~1- \alpha' P^2 =n
\label{onshephy}
\end{eqnarray}
These conditions characterizing the physical subspace were first found
by Del Giudice and Di Vecchia~\cite{DELGIU2} where the analysis 
described  here was done.

In order to find the physical subspace one starts writing the most
general on shell state contributing to  the residue of the pole at
level $n$ in Eq. (\ref{onshe67}). Then one imposes 
Eq.s (\ref{onshephy}) and determines the states that span
the physical subspace. Actually, among these states one finds 
also a set of zero norm states  that are physical and
spurious at the same time. Those states  are 
of the form given in Eq. (\ref{spu83}), but  also  
satisfy Eq.s (\ref{onshephy}). It is easy to see
that they are not really physical  because they are not contributing to the
residue of the pole at the level $n$. This follows from the form of
the unit operator given in the space of the physical states by:
\begin{eqnarray}
1 = \sum_{norm\, {\neq 0} } | \lambda , P \rangle \langle \lambda , P
| +\sum_{zero} \left[  | \lambda_0 , P \rangle \langle \mu_0 , P | + 
 | \mu_0 , P \rangle \langle \lambda_0 , P | \right]
\label{unit7}
\end{eqnarray} 
where $| \lambda_0 ,P \rangle$ is a zero norm physical and spurious
state and   $| \mu_0 ,P \rangle$ its  conjugate state. A conjugate
state of a zero norm state is obtained by changing the sign of the
oscillators with timelike direction. Since $|\lambda_0 , P \rangle $
is a spurious state when we insert the unit operator, given in 
Eq. (\ref{unit7}), in Eq. (\ref{factf1}) we see that the zero norm
states never contribute to the residue because their contribution is
annihilated either from the state $\langle p_{(1, M)}   |$ or from
the state $|p_{(M+1, N)}   \rangle$.  In conclusion, the physical 
subspace contains only the states in the first term in 
the r.h.s. of Eq. (\ref{unit7}). 

Let us analyze the first two excited levels. The first excited level
corresponds to a massless gauge field. It  is spanned by the states 
$\epsilon^{\mu} a_{1 \mu}^{\dagger} |0, P \rangle$. In this case the only
condition that we must impose is:
\begin{eqnarray}
L_1 \epsilon^{\mu} a_{1\mu}^{\dagger} |0, P \rangle=0 \Longrightarrow
P \cdot \epsilon =0
\label{L1}
\end{eqnarray} 
Choosing  a frame of reference where the momentum of the photon is given
by $P^{\mu} \equiv (P,0....0,P)$ , Eq. (\ref{L1}) implies that the only 
physical states are:
\begin{equation}
\label{2.6.15}
\epsilon^{i} a_{1i}^{+\dagger} |0, P \rangle  + 
\epsilon ( a_{1; 0}^{\dagger} - 
a_{1;d-1}^{\dagger} ) |0, P \rangle~~;~~ i =1 \dots d-2
\end{equation}
where $\epsilon^{i}$  and $\epsilon$  
are arbitrary parameters. The state in Eq.
(\ref{2.6.15}) is the most general state of the level $N=1$  satisfying the
conditions in Eq. (\ref{onshephy}). The first state in eq. (\ref{2.6.15}) has
positive norm, while the second one has zero norm that 
 is orthogonal to all other physical
states since it can be written as follows:
\begin{equation}
\label{2.6.16}
( a_{1;0}^{\dagger} - a_{1;D-1}^{\dagger}) |0, P \rangle =
L_{1}^{\dagger}|0, P \rangle
\end{equation}
in the frame of reference where $P^{\mu} \equiv ( P,...0, P)$. 
Because of the previous
property it is decoupled from the physical states together with its conjugate: 
\begin{equation}
\label{2.6.17}
( a_{1,0}^{\dagger} + a_{1,d-1}^{\dagger}) |0, P \rangle
\end{equation}
In conclusion, we are left only with the transverse $d-2$  states
corresponding to the physical degrees of freedom of a massless spin
$1$ state. At the next level $n=2$ the most general state is given by:
\begin{equation}
\label{2.6.18}
[ \alpha^{\mu \nu} a_{1,\mu}^{\dagger} a_{1,\nu}^{\dagger}  + 
\beta^{\mu} a_{2,\mu}^{\dagger} ] |0, P \rangle
\end{equation}
If we work in the center of mass frame where $P^{\mu} = ( M, \vec{0} )$
we get the following most general physical state:
\[
|Phys> = \alpha^{ij} [ a_{1,i}^{\dagger} a_{1,j}^{\dagger} - 
\frac{1}{(d-1)}\delta_{ij}\sum_{k=1}^{d-1}
a_{1,k}^{\dagger}a_{1,k}^{\dagger}]
 |0, P \rangle +
\]
\[
+ \beta^{i} [ a_{2,i}^{\dagger} + 
a_{1,0}^{\dagger} a_{1,i}^{\dagger}]|0, P >\rangle +
\]
\begin{equation}
\label{2.6.19}
+ \sum_{i=1}^{d-1} \alpha^{ii} \left[ \sum_{i=1}^{d-1} a_{1,i}^{\dagger} 
a_{1,i}^{\dagger} 
+ \frac{d-1}{5} ( a_{1,0}^{\dagger 2} - 2 a_{2,0}^{\dagger}) \right]
|0, P \rangle
\end{equation}
where the indices $i,j$ run over the $ d-1$ space components.
The first term in (\ref{2.6.19}) corresponds to a spin $2$  in $(d-1)$  
dimensional space and has a positive norm being made with space indices.
The second term has zero norm and is orthogonal to the other physical states
since it can be written as $L_{1}^{+} a_{1,i}^{+} |0, P \rangle$. 
Therefore it must be  eliminated
from the physical spectrum together with its conjugate, as explained above.
Finally, the last state in (\ref{2.6.19}) is spinless and has a norm given by:
\begin{equation}
\label{2.6.20}
2(d-1) ( 26-d)
\end{equation}
If $d< 26$  it corresponds to a physical spin zero particle with positive norm.
If $d>26$ it is a ghost. Finally, if $d=26$ it has a zero norm and is also
orthogonal to the other physical states since it can be written in the form:
\begin{equation}
\label{2.6.21}
( 2 L_{2}^{\dagger} + 3 L_{1}^{\dagger2} ) |0>
\end{equation}
It does not belong, therefore, to the physical spectrum.  The analysis of 
this level was done in Ref.~\cite{DELGIU} with $d=4$. This did not allow
the authors of Ref.~\cite{DELGIU}  to see that there was a critical dimension.

The analysis of the physical states can be easily extended~\cite{DELGIU2} 
to the Shapiro-Virasoro model. In this case the physical conditions 
given in Eq. (\ref{onshephy}) for the open string, become~\cite{DELGIU2}:
\begin{eqnarray}
L_m | \lambda, {\tilde{\lambda}} \rangle = 
{\tilde{L}}_m | \lambda, {\tilde{\lambda}} 
\rangle= (L_0 -1)   | \lambda, {\tilde{\lambda}} 
\rangle = ( {\tilde{L}}_0 -1)   | \lambda, {\tilde{\lambda}} \rangle=0 
\label{phy62}
\end{eqnarray}
for any  positive integer $m$.   It can be easily seen from the 
previous equations that the lowest state of the 
Shapiro-Virasoro model is the vacuum $| 0_a , 0_{\tilde{a}} , p \rangle $
corresponding to a tachyon with mass $\alpha ' p^2 = 4$, while 
the next level described by the state  
$ a^{\dagger}_{1\mu } {\tilde{a}}_{1\nu }^{\dagger} | 0_a , 
0_{\tilde{a}}, p  \rangle$
 contains massless states corresponding to the
graviton, a dilaton and a two-index antisymmetric tensor $B_{\mu \nu}$.

Having characterized the physical subspace
one can go on and construct a $N$-point  scattering amplitude
involving arbitrary  physical states. This was done by Campagna, 
Fubini, Napolitano and Sciuto~\cite{CAMPAGNA} 
where the vertex operator for an arbitrary physical
state was constructed in analogy with what has been done for the
ground tachyonic state. They associated to each physical state $
|\alpha, P \rangle$  a vertex operator $ V_{\alpha } (z, P )$ that is
a conformal field with conformal dimension equal to $1$:
\begin{eqnarray}
[ L_n , V_{\alpha} (z , p) ] =  \frac{d}{dz} \left( z^{n+1} V_{\alpha} 
(z , p) \right)
\label{commu6b}
\end{eqnarray}
and reproduces the corresponding state acting on the vacuum as
follows:
\begin{eqnarray}
\lim_{z \rightarrow 0} V_{\alpha} (z;p) | 0 ,0\rangle 
\equiv | \alpha ; p \rangle
~~;~~ \langle 0; 0 | \lim_{z \rightarrow \infty}  z^{2 }  
V_{\alpha} (z; p) = \langle \alpha, p |
\label{limi671b}
\end{eqnarray} 
It satisfies, in addition, the hermiticity relation:
\begin{eqnarray}
V_{\alpha}^{\dagger} (z, P ) = V_{\alpha} ( \frac{1}{z} , - P )
  (-1)^{\alpha ( - P^2 ) }
\label{hermi}
\end{eqnarray} 
An excited  vertex that will play an important role in the next
section is the one associated to the massless gauge field. It is given
by:
\begin{eqnarray}
V_{\epsilon} ( z , k ) \equiv 
\epsilon \cdot \frac{d Q (z)}{d z} {e}^{i k \cdot Q (z)}~~;~~ k \cdot
\epsilon = k^2 = 0
\label{phove}
\end{eqnarray}
Because of the last two  conditions in Eq. (\ref{phove}) the normal order
is not necessary. It is convenient to give the expression of 
$\frac{d Q (z)}{d z}$ in terms 
of the harmonic oscillators:
\begin{eqnarray}
P(z) \equiv  \frac{d Q (z)}{d z}  
= - i \sqrt{2 \alpha'} \sum_{n= -\infty}^{\infty} \alpha_n z^{-n -1}
 \label{dQ}
\end{eqnarray}
It is a conformal field with conformal dimension equal to $1$. The rescaled
oscillators $\alpha_n$ are given by:
\begin{eqnarray}
\alpha_n = \sqrt{n} a_{n}~~;~~\alpha_{-n} = \sqrt{n} a_{n}^{\dagger}
~~;~~n > 0~~;~~\alpha_0 = \sqrt{2 \alpha'} {\hat{p}}
\label{alphaos}
\end{eqnarray}

In terms of the vertex operators previously introduced 
the most general amplitude
involving arbitrary physical states is given by~\cite{CAMPAGNA}:
\begin{eqnarray}
( 2 \pi)^4 \delta ( \sum_{i=1}^{N}  p_i ) B_{N}^{ex} = 
\int_{-\infty}^{\infty} \frac{\prod_{1}^{N} d z_i  
\theta (z_i - z_{i+1}) }{dV_{abc}}
 \langle 0, 0 | \prod_{i=1}^{N} {{V_{\alpha_i}}} ( z_i , p_i)   |0, 0\rangle
\label{AN1b}
\end{eqnarray}

In the case of the Shapiro-Virasoro model the tachyon 
vertex operator is given in 
Eq. (\ref{veope}). By rewriting Eq. (\ref{QFV}) as follows:
\begin{eqnarray}
Q (z, {\bar{z}} )=  Q (z) + {\tilde{Q}} ({\bar{z}})
\label{}
\end{eqnarray}
where
\begin{eqnarray}
Q (z) = \frac{1}{2} \left[  {\hat{q}} - 2 \alpha' {\hat{p}} \log (z ) +
i  \sqrt{2 \alpha'}  \sum_{n=1}^{\infty} \frac{1}{\sqrt{n}} 
\left[ a_{n} z^{-n} - a_{n}^{\dagger} z^n \right] \right]
\label{Qz}
\end{eqnarray}
and 
\begin{eqnarray}
{\tilde{Q}} ({\bar{z}}) = \frac{1}{2} \left[  
{\hat{q}} - 2 \alpha' {\hat{p}} \log ( {\bar{z}}) 
+ i { \sqrt{2 \alpha'}} \sum_{n=1}^{\infty} \frac{1}{\sqrt{n}} 
\left[ {\tilde{a}}_{n} {\bar{z}}^{-n} - 
{\tilde{a}}_{n}^{\dagger} {\bar{z}}^n \right] \right]
\label{Qbarz}
\end{eqnarray}
we can write the tachyon vertex operator in the following way:
\begin{eqnarray}
V (z, {\bar{z}}, p ) = : e^{i p \cdot Q (z)}  e^{i p  
\cdot {\tilde{Q}} ({\bar{z}})} : 
\label{veclo}
\end{eqnarray}
This shows that the vertex operator corresponding to the 
tachyon of the Shapiro-Virasoro model can
be written as the product of two vertex operators 
corresponding each to the tachyon of 
the generalized Veneziano model. 

Analogously the 
vertex operator corresponding to an arbitrary   physical 
state of the Shapiro-Virasoro model can always be  written 
as a product of two vertex operators of 
the generalized Veneziano model: 
\begin{eqnarray}
V_{\alpha , \beta} ( z, {\bar{z}}, p ) = V_{\alpha} ( z, \frac{p}{2} ) 
V_{\beta} ( {\bar{z}}, \frac{p}{2} ) 
\label{veclo67}
\end{eqnarray}
The first one contains  only the oscillators $\alpha_n$, while  
the second one only the oscillators $ {\tilde{\alpha}}_n$. They both 
contain only 
half of the total momentum $p$ and 
the same zero modes ${\hat{p}}$ and ${\hat{q}}$. The two vertex operators
of the generalized Veneziano model are both conformal fields 
with conformal dimension equal to 1.  If they correspond to physical states 
at the level $2n$, they  satisfy the following 
relation $(n = {\tilde{n}})$:
\begin{eqnarray}
\alpha ' \frac{p^2}{4} + n =1  
\label{conf9s}
\end{eqnarray}
They lie on the following Regge trajectory:
\begin{eqnarray}
2 - \frac{\alpha'}{2} p^2 \equiv \alpha_{SV} ( - p^2 ) = 2 n
\label{traje}
\end{eqnarray}
as we have already seen by factorizing the amplitude in Eq. (\ref{fact38}).

\section{The DDF states and absence of ghosts}
\label{DDF}

In the previous section we have derived the equations that 
characterize the physical states and
their corresponding vertex operators. In this section we will
explicitly construct an infinite number of orthonormal 
physical states with positive norm.

The starting point is  the DDF operator introduced by 
Del Giudice, Di Vecchia and Fubini~\cite{DDF}
and defined in terms of the 
vertex operator corresponding to the massless gauge field introduced
in  eq. (\ref{phove}):
\begin{eqnarray}
A_{i, n} =  \frac{i}{\sqrt{2 \alpha'}} \oint_{0} {dz}
\epsilon_{i}^{\mu}  P_{\mu} ( z)  {e}^{i k \cdot Q (z)}
\label{trasta}
\end{eqnarray}
where the index {$ i$} runs over the $ d-2$ transverse directions, that
are orthogonal to the momentum $k$. We have also taken $\oint_0
\frac{dz}{z} =1$. Because of the  ${\log z}$ term
appearing in the zero mode part of the exponential,  the integral in
Eq. (\ref{trasta}), that is performed around the origin {$ z=0$}, 
is well defined only if we constrain the momentum of the state, on 
which $A_{i,n}$  acts, to satisfy the relation:
\begin{eqnarray}
2 \alpha' p \cdot k =  n
\label{cond7}
\end{eqnarray}
where {$ n$} is a non-vanishing  integer. 

The operator in Eq. (\ref{trasta}) will generate physical states 
because it commutes with the gauge operators $L_m$:
\begin{equation}
[ L_m , A_{n;i}] =0
\label{com87}
\end{equation}
since the  vertex operator transforms as a primary field with
conformal dimension equal to $1$ as it follows from
Eq. (\ref{commu6b}).  

On the other hand it also satisfies the algebra of the harmonic
oscillator as we are now going to show. From Eq. (\ref{trasta})
we get:
\begin{equation}
[ A_{n,i} , A_{m, j} ] = -\frac{1}{2 \alpha'} 
\oint_{0} d \zeta \oint_{\zeta} dz  \epsilon_{i} 
\cdot  P( z)   {e}^{i k \cdot Q (\zeta )} \epsilon_{j} \cdot 
P (\zeta ) {e}^{i k' \cdot Q ( \zeta )}
\label{commu46}
\end{equation}
where
\begin{equation}
2 \alpha' p \cdot k =  n~~;~~ 2 \alpha' p \cdot k' = m
\label{con45}
\end{equation}
and $ k $  and $ k'  $  are supposed to be in the same direction,
namely 
\begin{equation}
k_{\mu} =  n {\hat{k}}_{\mu} ~~~;~~~ k_{\mu}' =  m {\hat{k}}_{\mu}
\label{kk'}
\end{equation}
with
\begin{equation}
2 \alpha' p \cdot {\hat{k}} =1
\label{con62b}
\end{equation}
Finally the polarizations are normalized as:
\begin{eqnarray}
\epsilon_i \cdot \epsilon_j = \delta_{ij}
\label{nor56}
\end{eqnarray}
Since $ {\hat{k}} \cdot \epsilon_{i} =  {\hat{k}} \cdot \epsilon_{j} =
{\hat{k}}^2 =0$  a singularity for $ z= \zeta  $  can appear only from the
contraction of the two   terms $ P (\zeta ) $ and $ P ( (z)$ 
that is given by:
\begin{equation}
\langle 0,0 |  \epsilon_{i} \cdot  P (z)  
\epsilon_{j} \cdot P (\zeta ) |0, 0  \rangle = 
 -\frac{2 \alpha' \delta_{ij}}{(z - \zeta)^2}
\label{contrab}
\end{equation}
Inserting  it  in Eq. (\ref{commu46}) we get:
\[
[ A_{n,i} , A_{m, j} ] = \delta_{ij} in \oint_{0} d \zeta {\hat{k}} \cdot P 
(\zeta )  {e}^{- i (n+m) ) {\hat{k}} 
\cdot Q (\zeta) }=
\]
\begin{eqnarray}
 = in \delta_{ij}  \delta_{n+m;0}
\oint_0 d \zeta {\hat{k}} \cdot P (\zeta)  
\label{commu56b}
\end{eqnarray}
where we have used the fact that 
the integrand  is a total derivative and therefore one gets  a
vanishing contribution unless $n+m =0$. If $n+m =0 $  from Eq.s (\ref{dQ}) and
(\ref{con62b})  we get:
\begin{eqnarray}
[ A_{n,i} , A_{m, j} ] =  n \delta_{ij} \delta_{n+m;0}~~;~~i,j = 1 \dots d-2
\label{commu71b}
\end{eqnarray}
 Eq. (\ref{commu71b}) shows that
the DDF operators satisfy the harmonic oscillator algebra.

In terms of this infinite set of  transverse oscillators 
we can construct an orthonormal  set of states:
\begin{eqnarray}
| i_1 , N_{1}; i_2 , N_2 ; \dots i_m , N_m \rangle = \prod_{h}
  \frac{1}{\sqrt{\lambda_h ! }} \prod_{k=1}^{m}
  \frac{A_{i_k , - N_k}}{\sqrt{N_k}} | 0, p \rangle
\label{ortho6}  
\end{eqnarray}
where $ \lambda_h  $  is the multiplicity of the operator $ A_{i_h
  ,-N_h}$  in the product in Eq. (\ref{ortho6}) and 
the momentum of the state in Eq. (\ref{ortho6}) is given by
\begin{eqnarray}
P = p + \sum_{i=1}^{m} {\hat{k}} N_i
\label{totmo}
\end{eqnarray}
They were constructed in four dimensions where they were not a
complete system of states~\footnote{Because of this  Fubini did not want 
to publish our result, but then he went to a meeting in Israel in spring 1971
giving a talk on our work where he found that the audience 
was very interested in our result and when he came back 
to MIT we decided to publish our result.}  
and it took some time to realize that in
fact they were a complete system of states if 
$d=26$~\cite{BROWER,GT}~\footnote{I still remember 
Charles Thorn coming into my office at Cern and telling me: 
Paolo, do you know that your DDF states are complete
if $d=26  ?$  I quickly redid the analysis  done in Ref.~\cite{DELGIU} with
an arbitrary value of the space-time dimension obtaining    
Eq.s (\ref{2.6.19}) and (\ref{2.6.20}) that show that the spinless 
state at the level $\alpha (s) =2$ is decoupled if $d=26$. 
I strongly regretted not to have used an arbitrary space-time 
dimension d in the analysis of Ref.~\cite{DELGIU} . }.
Brower~\cite{BROWER} and  Goddard and Thorn~\cite{GT} showed also that
the dual resonance model was ghost free for any dimension
$d \leq 26$. In $d=26$ this follows from the fact that the DDF operators  
obviously span a positive definite Hilbert space (See Eq. (\ref{commu71b})). 
For $ d < 26$ there are extra states
called Brower states~\cite{BROWER}. The first of these states is the
last state in Eq. (\ref{2.6.19}) that becomes a zero norm state for $d=26$. 
But also for $d <26$ there is no negative norm state among the 
physical states. The proof of the no-ghost theorem in the case $\alpha_0 =1$
is a very important step because it shows that the dual resonance model 
constructed generalizing the four-point 
Veneziano formula, is a fully consistent
quantum-relativistic theory!  This is not quite 
true because, when the intercept
$\alpha_0 =1$, the lowest state of the spectrum corresponding to the pole
in the $N$-point amplitude for $\alpha (s) =0$, is a tachyon with mass 
$ m^2 = - \frac{1}{\alpha'}$.  A lot of effort was then made to construct a 
model without tachyon and with a meson spectrum consistent 
with the experimental data. The only reasonably consistent models that
came out from these attempts, 
were the Neveu-Schwarz~\cite{NS} for mesons and 
the Ramond model~\cite{RAMOND} for
fermions that only later were recognized to be part of a unique model
that  nowadays is  called  the Neveu-Schwarz-Ramond model. But this model
was not really more consistent than the original dual resonance model
because it still had a tachyon with mass $m^2 = -\frac{1}{2 \alpha'}$. The
tachyon was eliminated from the spectrum only in 1976 
through the GSO projection
proposed by Gliozzi, Scherk and Olive~\cite{GSO}.

Having realized that, at least for the critical value of the space-time 
dimension $d=26$, the physical states are described by the DDF states
having only $d-2= 24$ independent components,  open the way 
to Brink and Nielsen~\cite{BNIELSEN} to compute the 
value $\alpha_0 =1$ of the 
Regge trajectory with a very physical argument. They related the intercept
of the Regge trajectory to the zero point energy of a system with
an infinite number of oscillators having only $d-2$ independent components:
\begin{eqnarray}
\alpha_0 = - \frac{d-2}{2} \sum_{n=1}^{\infty} n
\label{zeropo}
\end{eqnarray}
This quantity is obviously infinite and, in order to make sense of it, they 
introduced a cutoff on the frequencies of the harmonic oscillators 
obtaining an infinite term that they eliminated by renormalizing the 
speed of light and a finite universal constant term  that gave the intercept
of the Regge trajectory. Instead of following their original approach we
discuss here an alternative approach due  to  Gliozzi~\cite{GLIOZZI2}
that uses the $\zeta$-function regularization. He rewrites Eq. (\ref{zeropo})
as follows:
\begin{eqnarray}
\alpha_0 = - \frac{d-2}{2} \sum_{n=1}^{\infty} n =  - \frac{d-2}{2} 
\lim_{s\rightarrow -1} \sum_{n=1}^{\infty}  n^{-s}  =  - 
\frac{d-2}{2} \zeta_{R} (-1 ) =  1  
\label{zeropo1}
\end{eqnarray}
where  in the last equation we have used the identity 
$\zeta_{R} (-1 ) = - \frac{1}{12}$ 
and we have put $d=26$. Since the Shapiro-Virasoro model has 
two sets of transverse harmonic oscillators it is obvious 
that its intercept is 
twice that of the generalized Veneziano model.

Using the rules discussed in the previous section we can construct the
vertex operator corresponding to the state in Eq. (\ref{ortho6}). It
is given by:
\begin{eqnarray}
V_{(i; N_{i})} ( z, P ) = \prod_{i=1}^{m} \oint_{z} d z_i \epsilon_{i}
\cdot P (z_i ) {e}^{i N_{i} {\hat{k}} \cdot Q( z_i ) } :{e}^{i p
  \cdot Q (z) } :
\label{ver93}
\end{eqnarray}
where the integral on the variable $z_i$ is evaluated along a curve
of the complex plane $z_i$ containing the point $z$. The singularity
of the integrand for $z_i =z$ is a pole provided that the following
condition is satisfied.
\begin{eqnarray}
2 \alpha' p \cdot {\hat{k}} =1
\label{con62c}
\end{eqnarray}
The last vertex in Eq. (\ref{ver93}) is the vertex operator
corresponding to the ground tachyonic state given in
Eq. (\ref{vertope}) with $\alpha ' p^2 =1$.

Using the general form of the vertex one can compute the three-point
amplitude involving three arbitrary DDF vertex operators. This
calculation has been performed in Ref.~\cite{ADDF} and  since the vertex 
operators are conformal fields with dimension equal to $1$ one gets:
\[
\langle 0,0 | V_{(i^{(1)}_{k_1} ; N_{k_1}^{(1)} )   } ( z_1 , P_1 ) 
V_{(i^{(2)}_{k_2}; N_{k^{(2)}}^{(2)} )} ( z_2 , P_2 ) 
 V_{(i^{(3)}_{k_3} ; N_{k^{(3)}}^{(3)} )} ( z_3 , P_3 ) | 0, 0 \rangle = 
\]
\begin{eqnarray}
= \frac{C_{123}}{(z_1 - z_2 ) (z_1 - z_3 )( z_2 - z_3)}
\label{3ver}
\end{eqnarray}
where the explicit form of the coefficient $C_{123}$ is given by:
\[
C_{123} = {}_{1} \langle 0, 0| {}_{2} \langle 0, 0| {}_{3} \langle 0,
0|
{e}^{\frac{1}{2} 
\sum_{r.s=1}^{3} \sum_{n,m =1}^{\infty}
A^{(r)}_{-n ; i} N^{rs}_{nm}  A^{(s)}_{-m ; i} +
\sum_{i=1}^{3} P_i \cdot \sum_{n=1}^{\infty} A^{(r)}_{-n ;i} } \times
\]
\begin{eqnarray}
\times {e}^{\tau_0 \sum_{r=1}^{3} (\alpha ' \Pi_{r}^{2} - 1)} |
N_{k_1}^{(1)} , i^{(1)}_{k_1}  
\rangle_{1}  | N_{k_2}^{(2)} , i^{(2)}_{k_2} \rangle_{2}
| N_{k_3}^{(3)} , i^{(3)}_{k_3} \rangle_{3}
\label{ver52b}
\end{eqnarray}
where
\begin{eqnarray}
N_{nm}^{rs} = - N_{n}^{r} N^{s}_{m} 
\frac{nm \alpha_1 \alpha_2 \alpha_3 }{n \alpha_s + m \alpha_r}~~~;~~~
N_{n}^{r} = \frac{\Gamma ( - n \frac{\alpha_{r+1} }{\alpha_r})}{\alpha_r
  n! \Gamma ( 1 - n  \frac{\alpha_{r+1}}{\alpha_r} -n)}
\label{neucoe}
\end{eqnarray}
with
\begin{eqnarray}
\Pi=  P_{r+1} \alpha_r -  P_r \alpha_{r+1}~~~;~~~r=1,2,3
\label{Pi}
\end{eqnarray}
$\Pi$ is independent on the value of $r$ 
chosen as a consequence of the equations:
\begin{eqnarray}
\sum_{r=1}^{3}  \alpha_r =  \sum_{r=1}^{3} P_r  =0
\label{conse42}
\end{eqnarray}

\section{The zero slope limit}
\label{zeroslope}

In the introduction we have seen that the dual resonance model 
has been constructed using rules that are different from those 
used in field theory. For instance, we have seen that planar duality 
implies that the amplitude corresponding to a certain duality
diagram, contains   poles in both s and t channels, while 
the amplitude corresponding to a  Feynman  diagram in field 
theory contains only a pole in one of the two channels.  Furthermore,
the scattering amplitude in the dual resonance model contains 
an infinite number of resonant states that, at high energy,  average out
to give  Regge behaviour. Also this property  is not observed in
field theory. The question that was natural to ask, was then: 
is there any relation  between the dual resonance model and field theory?
It turned out, to the surprise of many, that the dual resonance model
was not in contradiction with field theory, but was instead an extension 
of a certain number of field theories. We will
see that the limit in which a field theory is obtained from the dual 
resonance model corresponds to taking the slope of the Regge 
trajectory $\alpha'$ to zero. 

Let us consider the scattering amplitude of four ground state particles 
in Eq. (\ref{astu}) that we rewrite here with the correct normalization factor:
\begin{eqnarray}
A (s, t, u) = C_0 N_{0}^{4} \left( A(s,t) + A (s,u) + A (t,u)  \right)
\label{astub}
\end{eqnarray}
where
\begin{eqnarray}
N_0 = \sqrt{2} g (2 \alpha' )^{\frac{d-2}{4}}
\label{N0b}
\end{eqnarray}
is the correct normalization factor for each external leg, g is the 
dimensionless open string coupling constant that we have
constantly ignored in the previous sections and $C_0$ is 
determined by the following relation:
\begin{eqnarray}
C_0 N_{0}^{2} \alpha' = 1
\label{facto45}
\end{eqnarray}
that is obtained by requiring the factorization of the amplitude at the
pole corresponding to the ground state particle whose mass is given
in Eq. (\ref{m2}).  Using Eq. (\ref{m2}) in order to rewrite the intercept
of the Regge trajectory in terms of the mass of the ground state 
particle $m^2$ and  the following relation satisfied by 
the $\Gamma$- function:
\begin{eqnarray}
\Gamma (1 +z ) = z \Gamma (z)
\label{gam78}
\end{eqnarray}
we can easily perform the limit for $\alpha' \rightarrow 0$ of $A (s, t)$
obtaining:
\begin{eqnarray}
\lim_{\alpha ' \rightarrow 0} A (s, t)  = 
\frac{1}{\alpha' } \left[  \frac{1}{m^2 -s} 
+   \frac{1}{m^2 -s}  \right] 
\label{lim82}
\end{eqnarray}
Performing the same limit on the other two planar amplitudes we
get the following expression for the total amplitude in Eq. (\ref{astub}):
\begin{eqnarray}
\lim_{\alpha ' \rightarrow 0} A (s, t, u)  = 
\left[  \sqrt{2} g (2 \alpha' )^{\frac{d-2}{4}} 
\right]^2  \frac{2}{(\alpha' )^2}
\left[  \frac{1}{m^2 -s}  +   \frac{1}{m^2 -s}  + \frac{1}{m^2 -u} \right]   
\label{lim83}
\end{eqnarray}
By introducing the coupling constant:
\begin{eqnarray}
g_3 = 4 g  (2 \alpha' )^{\frac{d-6}{4}}
\label{g3}
\end{eqnarray}
Eq. (\ref{lim83}) becomes
\begin{eqnarray}
\lim_{\alpha ' \rightarrow 0} A (s, t, u)  =  g_{3}^{2} 
\left[  \frac{1}{m^2 -s}  +   \frac{1}{m^2 -s}  + \frac{1}{m^2 -u} \right]  
\label{lim89}
\end{eqnarray}
that is equal to the sum of the tree diagrams for the scattering of 
four particles with mass $m$ of $\Phi^3$ theory with coupling 
constant equal to $g_3$.  We have shown that, 
by keeping $g_3 $ fixed in the limit $\alpha' \rightarrow 0$, 
the scattering amplitude of four ground state particles of the dual
resonance model is
equal to the  tree diagrams of $\Phi^3$ theory.  This 
proof can be extended to the scattering of $N$ ground state particles
recovering also in this case the tree diagrams of $\Phi^3$ theory. It is also
valid for loop diagrams that we will discuss in the next section.
In conclusion, the dual resonance model reduces in the zero slope
limit to  $\Phi^3$ theory.  The proof that we have presented here is 
due to J. Scherk~\cite{scherk}~\footnote{See also Ref.~\cite{naka}.}

A more interesting case to study is the one with intercept $\alpha_0 =1$.
We will see that, in this case, one will obtain the tree diagrams of
Yang-Mills theory, as shown by   Neveu and
Scherk~\cite{NEVEUS}~\footnote{See also Ref.~\cite{NEVEUG}.}.

Let us consider the three-point amplitude involving three massless
gauge particles described by the vertex operator in
Eq. (\ref{phove}). It is given by the sum of two planar diagrams. The
first one corresponding to the ordering $(123)$ is given by:
\begin{eqnarray}
C_0 N_{0}^{3} i^3 Tr \left(\lambda^{a_1} \lambda^{a_2} \lambda^{a_3}  \right)
\frac{\langle 0 , 0| V_{\epsilon_1} (z_1 , p_1 ) V_{\epsilon_2} (z_2 , p_2 )
 V_{\epsilon_3} (z_3 , p_3 ) |0, 0 \rangle}{\left[(z_1 - z_2 )(z_2 -
   z_3) (z_1 - z_3 ) \right]^{-1}}
\label{3poi}
\end{eqnarray}
Using momentum conservation $p_1 + p_2 + p_3 =0$ and the mass shell conditions 
$p_{i}^{2} = p_i \cdot \epsilon_i =0$ one can rewrite the previous
equation as follows:
\[
C_0 N_{0}^{3} Tr( \lambda^{a_1} \lambda^{a_2} \lambda^{a_3})  \sqrt{2 \alpha'}
\times
\]
\begin{eqnarray}
\times \left[ (\epsilon_1 \cdot \epsilon_2) ( p_1 \cdot \epsilon_3 ) + 
( \epsilon_1 \cdot \epsilon_3 ) (p_3 \cdot \epsilon_2 ) + 
(\epsilon_2 \cdot \epsilon_3 ) ( p_2 \cdot \epsilon_1)  \right]
\label{3gluon}
\end{eqnarray}
The second contribution comes from the ordering $132$ that can 
be obtained from the previous one by the substitution
\begin{eqnarray}
Tr( \lambda^{a_1} \lambda^{a_2} \lambda^{a_3}) \rightarrow 
- Tr( \lambda^{a_1} \lambda^{a_3} \lambda^{a_2})
\label{anticy4}
\end{eqnarray}
Summing the two contributions one gets
\[
C_0 N_{o}^{3} Tr( \lambda^{a_1} [\lambda^{a_2}, \lambda^{a_3}])  
\sqrt{2 \alpha'} \times
\]
\begin{eqnarray}
\times
\left[ (\epsilon_1 \cdot \epsilon_2) ( p_1 \cdot \epsilon_3 ) + 
( \epsilon_1 \cdot \epsilon_3 ) (p_3 \cdot \epsilon_2 ) + 
(\epsilon_2 \cdot \epsilon_3 ) ( p_2 \cdot \epsilon_1)  \right]
\label{3gluon43}
\end{eqnarray}
The factor
\begin{eqnarray}
N_0 = 2 g  (2 \alpha')^{(d-2)/4}
\label{N0}
\end{eqnarray}
is the correct normalization factor for each vertex operator if  we
normalize the generators of the Chan-Paton group as follows:
\begin{eqnarray}
Tr \left(  \lambda^i \lambda^j \right) = \frac{1}{2} \delta^{ij}
\label{cpfa}
\end{eqnarray}
It is related to $C_0$ through the relation~\footnote{The determination
of  the previous normalization factors can be found in the
Appendix of Ref.~\cite{DLMMR}.}:
\begin{eqnarray}
C_0 N_{o}^{2} \alpha ' =2
\label{facto83}
\end{eqnarray}
$g$ is the dimensionless open string coupling constant.
Notice that Eq.s (\ref{N0}) and (\ref{facto83}) differ from Eq.s
(\ref{N0b}) and (\ref{facto45}) because of the presence of the
Chan-Paton factors that we did not include in the case of $\Phi^3$ theory.

By using the commutation relations:
\begin{equation}
[ \lambda^a , \lambda^b ] = i f^{abc} \lambda^c
\label{commu76}
\end{equation}
and the previous normalization factors we get for the three-gluon amplitude:
\[
i g_{YM} f^{a_1 a_2 a_3 } \left[ (\epsilon_1 \cdot \epsilon_2) 
( (p_1 - p_2) \cdot \epsilon_3  \right. +
\]  
\begin{eqnarray}
 \left. 
+ 
( \epsilon_1 \cdot \epsilon_3 ) ( (p_3 - p_1) \cdot \epsilon_2 ) + 
(\epsilon_2 \cdot \epsilon_3 ) ( (p_2 - p_3) \cdot \epsilon_1)  \right]
\label{3gluonfin}
\end{eqnarray}
that is equal to the $3$-gluon vertex that one obtains from the Yang-Mills
action
\begin{eqnarray}
L_{YM} = - \frac{1}{4 } F_{\alpha \beta}^{a} F^{\alpha \beta}_{a}~~~,~~~
F_{\alpha \beta}^{a} = \partial_{\alpha} A_{\beta}^{a} - \partial_{\beta} 
A_{\alpha}^{a} + g_{YM} f^{abc} A_{\alpha}^{b} A_{\beta}^{c} 
\label{gaugela34}
\end{eqnarray}
where
\begin{eqnarray}
g_{YM} = 2 g ( 2 \alpha ' )^{\frac{d-4}{4}}
\label{gYM}
\end{eqnarray}
The previous procedure can be extended to the scattering of N gluons
finding the same result that one gets from the tree diagrams of
Yang-Mills theory. In the next section, we will discuss the loop
diagrams. Also, in this case one finds that the h-loop diagrams
involving N external gluons reproduces in the zero slope limit the sum
of the h-loop diagrams with N external gluons of Yang-Mills theory.

We conclude this section mentioning that one can also take the zero
slope limit of a scattering amplitude involving three and four
gravitons obtaining agreement with what one gets from the Einstein
Lagrangian of general relativity. This has been shown by
Yoneya~\cite{YONEYA}.

\section{Loop diagrams}
\label{loop}

The $N$-point amplitude previously 
constructed satisfies all the 
axioms of S-matrix theory except unitarity because its only 
singularities are simple poles 
corresponding to zero width resonances lying on the real axis of the
Mandelstam variables and 
does not contain the various cuts required 
by unitarity~\cite{ESSE}.  In order to eliminate
this problem  it was proposed already
in the early days of dual theories to assume, in analogy with what happens 
for instance in perturbative field  theory,  that the 
$N$-point amplitude was only the lowest 
order (the tree diagram) of a perturbative expansion 
and, in order to implement unitarity, it was 
necessary to include loop diagrams.  Then, the one-loop diagrams
were constructed from the propagator and vertices that we have introduced in
the previous sections~\cite{KSV}. 
The planar one-loop amplitude with $M$ external particles
was computed by  starting from a $(M+2)$-point tree amplitude and then
by sewing two external legs together after the insertion of a propagator $D$ 
given in Eq. (\ref{proab}). In this way one gets:
\begin{eqnarray}
\int \frac{d^{d} P}{(2 \alpha')^{d/2} 
(2 \pi)^d} \sum_{\lambda} \langle P, \lambda | 
V (1, p_1 ) D  V (1, p_2) \dots V (1, p_N ) D | P, \lambda \rangle
\label{onelo}
\end{eqnarray}
where the sum over $\lambda$ corresponds to the trace in the space 
of  the 
harmonic oscillators and the integral in   $d^d  P$  
corresponds to integrate  over the momentum circulating in the loop.  
The previous expression for the one-loop amplitude cannot be 
quite correct because all states of the space generated by the 
oscillators in Eq. (\ref{osci}) are circulating in the loop, while we know
that we  should include  only the  physical ones. 
This was achieved first by cancelling by hand the time 
and one of the space components of the harmonic oscillators reducing the
degrees of freedom of each oscillator from $d$ to $d-2$ as suggested by
the DDF operators at least for  $d=26$. This procedure was then shown
to be correct by Brink and Olive~\cite{BO}. 
They  constructed the operator that projects over 
the physical states and, by inserting it in the loop,  showed 
that the reduction of  the degrees of freedom of the oscillators from $d$ to
$d-2$ was indeed correct. This was, at that time, the only procedure available
to let only the physical states circulate in the loop 
because the BRST procedure
was discovered a bit  later also in the framework of the gauge field
theories!  

To be more explicit let us compute the trace in Eq. (\ref{onelo})
adding  also the  Chan-Paton factor. We get:
\[
(2\pi)^{d} \delta^{(d)} \left( \sum_{i=1}^{M} p_i \right) 
\frac{N Tr ( \lambda^{a_1} \dots \lambda^{a_M} ) }{ ( 8 \pi^2 \alpha' )^{d/2} }
N_{0}^{M}   \int_{0}^{\infty} \frac{d \tau}{ \tau^{d/2 +1} }
[f_{1} (k ) ]^{2-d}  k^{\frac{d-26}{12}}   (2 \pi )^{M} \times 
\]
\begin{eqnarray}
\times \int_{0}^{1} d \nu_{M} \int_{0}^{\nu_{M}} d
  \nu_{M-1} \dots  \int_{0}^{\nu_{3}} d \nu_{2} \,\,\tau^M
 \prod_{i < j} \left[e^{G (\nu_{ji}) } \right]^{2 \alpha' p_i \cdot p_j}
 ; k \equiv  e^{- \pi \tau}
\label{planar}
\end{eqnarray}
where $\nu_{ji} \equiv \nu_j - \nu_i$,
\begin{eqnarray}
G (\nu ) = \log \left( i  e^{-\pi \nu^{2}   \tau}  
\frac{\Theta_1 ( i \nu \tau | i \tau)}{f_{1}^{3} (k)}  \right)~~;~~
f_{1} (k) =   k^{1/12} \prod_{n=1}^{\infty} ( 1 - k^{2n} )
\label{green5}
\end{eqnarray}
and 
\begin{eqnarray}
\Theta_1 ( \nu | i \tau ) = -2 k^{1/4} \sin \pi \nu \prod_{n=1}^{\infty} 
{\left(1 - e^{2i \pi \nu} k^{2n} \right) 
\left(1 - e^{-2i \pi \nu} k^{2n} \right) }{ (1- k^{2n} )}
\label{theta1}
\end{eqnarray}
Finally the normalization factor $N_0$ is given in Eq. (\ref{N0}). 
We have performed the calculation for an arbitrary value of the space-time 
dimension d. However, in this way one gets also the extra factor of 
$k^{\frac{d-26}{12}}$ appearing in the first 
line of Eq. (\ref{planar})  that implies  that our calculation 
is actually  only  consistent if $d=26$. In fact, the presence of this 
factor does not
allow one to rewrite the amplitude, originally  obtained in the Reggeon sector, 
in the Pomeron  sector as explained below. In the following we neglect this
extra factor,  implicitly assuming that   $d=26$,  but, on the other hand, still
keeping  an arbitrary $d$.

Using the relations:
\begin{eqnarray}
f_1 (k) = \sqrt{t} f_{1} (q)~~;~~\Theta_1 (i \nu \tau | i \tau ) = 
i \Theta_1 ( \nu | i t ) t^{1/2} e^{\pi \nu^2 /t}  
\label{trasfo76}
\end{eqnarray}
where $ t = \frac{1}{\tau}$ and $ q \equiv e^{-\pi t}$, we 
can rewrite the one-loop planar diagram in the Pomeron channel.
We get:
\[
(2\pi)^{d} \delta^{(d)} \left( \sum_{i=1}^{M} p_i \right)
\frac{N Tr ( \lambda^{a_1} \dots \lambda^{a_M} )}{( 8 \pi^2 \alpha')^{d/2}}
N_{0}^{M}   \int_{0}^{\infty}  dt
[f_{1} (q ) ]^{2-d}   (2 \pi)^M \times 
\]
\begin{eqnarray}
\times \int_{0}^{1} d \nu_{M} \int_{0}^{\nu_{M}} d
  \nu_{M-1} \dots  \int_{0}^{\nu_{3}} d \nu_{2}  
 \prod_{i < j} \left[   - \frac{\Theta_1 ( \nu_{ji} | it)}{f_{1}^{3} (q)}
  \right]^{2 \alpha' p_i \cdot p_j}
\label{clo84}
\end{eqnarray}
Notice that, by factorizing the planar loop in the Pomeron channel, 
one  constructed for the first time what we
now call the boundary state~\cite{POME}~\footnote{See also the first
  paper in Ref.~\cite{ADE2}.}. This can be easily seen in
the way that we are now going to describe. First of all, notice that
the last quantity in Eq. (\ref{clo84}) can be written as follows:
\[
 \prod_{i < j}
\left[   - \frac{\Theta_1 ( \nu_{ji} | it)}{f_{1}^{3} (q)}
  \right]^{2 \alpha' p_i \cdot p_j} =  
\]
\begin{eqnarray}
=
\prod_{i < j} \left[ 
- 2 \sin (\pi \nu_{ji}) \prod_{n=1}^{\infty} \frac{
\left( 1 - q^{2n} e^{2 \pi i \nu_{ji}} \right)
\left( 1 - q^{2n} e^{- 2 \pi i \nu_{ji}} \right) }{(1 - q^{2n} )^2}
\right]^{2 \alpha' p_i \cdot p_j}
\label{form89}
\end{eqnarray}
This equation  can be rewritten as
follows:
\begin{eqnarray}
\frac{Tr \left( \langle p=0 | q^{2R} \prod_{i=1}^{M} 
: e^{ i p_i \cdot Q ( e^{2
      i \pi \nu_i}) } :   | p=0 \rangle \right)  i^M
}{Tr \left( \langle p=0 | q^{2N} | p=0 \rangle \right) }~;~ R =
      \sum_{n=1}^{\infty} n a^{\dagger}_{n} \cdot a_{n}
\label{formu}
\end{eqnarray}
where the trace is taken only over the non-zero modes and momentum
conservation has been used. It must also be stressed that the normal
ordering of the vertex operators in the previous equation is such that
the zero modes are taken to be both in the same exponential instead of
being ordered as in Eq. (\ref{vertope}). By bringing all
annihilation operators on the left of the creation ones, from
the expression in Eq. (\ref{formu}) one gets 
$(z_i \equiv e^{2 \pi i \nu_i})$:
\[
(2 \pi )^d \delta^{(d)} \left( \sum_{i=1}^{\infty}  p_i \right)
\prod_{i < j} ( -2 \sin \pi \nu_{ji} )^{2 \alpha' p_i \cdot p_j}
\times
\]
\begin{eqnarray}
\times
\frac{\prod_{i.j} \prod_{n=1}^{\infty} Tr \left(q^{2n a_{n}^{\dagger} \cdot
    a_n }
e^{\sqrt{2 \alpha'} p_j \cdot 
\frac{a_{n}^{\dagger} }{\sqrt{n}} z_{j}^{n} } e^{- \sqrt{2 \alpha'} p_i \cdot 
\frac{a_{n} }{\sqrt{n}} z_{i}^{-n} } \right)}{ Tr 
\left( \langle p=0 | q^{2N} | p=0 \rangle \right)}
\label{formu72}
\end{eqnarray}
The trace can be computed by using the completeness relation involving
coherent states $ | f \rangle = e^{f a^{\dagger}} | 0 \rangle$:
\begin{eqnarray}
 \int \frac{d^2 f}{\pi} e^{- | f|^2}  | f \rangle \langle f | = 1 
\label{identi}
\end{eqnarray}
Inserting the previous identity operator in Eq. (\ref{formu72}) one
gets after some calculation:
\[
(2 \pi )^d \delta^{(d)} \left( \sum_{i=1}^{\infty}  p_i \right)
\prod_{i < j} ( -2 \sin \pi \nu_{ji} )^{2 \alpha' p_i \cdot p_j} \times
\]
\begin{eqnarray}
\times \prod_{i.j=1}^{M} \prod_{n=1}^{\infty} e^{ - 2 \alpha' p_i \cdot p_j
e^{ 2 \pi i n \nu_{ji}} \frac{q^{2n}}{{n} (1 - q^{2n})} }
\label{formu73}
\end{eqnarray}
Expanding the denominator in the last exponent and performing the sum
over $n$ one gets:
\[
(2 \pi )^d \delta^{(d)} \left( \sum_{i=1}^{\infty}  p_i \right)
\prod_{i < j} ( -2 \sin \pi \nu_{ji} )^{2 \alpha' p_i \cdot p_j} \times
\]
\begin{eqnarray}
\times
 \prod_{i.j}  e^{ 2 \alpha' p_i \cdot p_j
   \sum_{m=0}^{\infty} \log \left(1 - e^{2 \pi i \nu_{ji}} q^{2 (m+1)} 
\right) } 
\label{formu75}
\end{eqnarray}
that is equal to the last line of Eq. (\ref{form89}) apart from the
$\delta$-function for momentum conservation. In conclusion, we have
shown that Eq.s (\ref{form89}) and (\ref{formu}) are equal.

Using Eq. (\ref{form89})
we can rewrite Eq. (\ref{clo84}) as follows:
\[
\frac{N  N_{0}^{M} 
Tr ( \lambda^{a_1} \dots \lambda^{a_M} )}{( 8 \pi^2 \alpha')^{d/2}}
  \int_{0}^{\infty}  dt
[f_{1} (q ) ]^{2-d} (2 \pi i )^M 
\int_{0}^{1} d \nu_{M} \int_{0}^{\nu_{M}} d
  \nu_{M-1} \dots    
\]
\begin{eqnarray}
\dots \int_{0}^{\nu_{3}} d \nu_{2} \frac{ \sum_{\lambda} 
 \langle p=0 , \lambda | q^{2R} \prod_{i=1}^{M} : e^{ i p_i \cdot Q ( e^{2
      i \pi \nu_i}) } :   | p=0, \lambda \rangle   
}{\sum_{\lambda}  \langle p=0, \lambda | q^{2N} | p=0, \lambda 
\rangle }
\label{clo89}
\end{eqnarray}
where the sum over any state $ |\lambda \rangle$ corresponds to 
taking the trace
over the non-zero modes. If $d=26$ we can rewrite Eq. (\ref{clo89}) in
a simpler form:
\[
\frac{N  N_{0}^{M} 
Tr ( \lambda^{a_1} \dots \lambda^{a_M} )}{( 8 \pi^2 \alpha')^{d/2}}
  \int_{0}^{\infty}  dt\,                  (2 \pi i )^M 
\int_{0}^{1} d \nu_{M} \int_{0}^{\nu_{M}} d
  \nu_{M-1} 
\dots \int_{0}^{\nu_{3}} d \nu_{2} \times
\]
\begin{eqnarray}
\times \sum_{\lambda} 
 \langle p=0 , \lambda | q^{2R-2} \prod_{i=1}^{M} : e^{ i p_i \cdot Q ( e^{2
      i \pi \nu_i}) } :   | p=0, \lambda \rangle 
\label{clo86}
\end{eqnarray}  
The previous equation contains the factor $\int dt q^{2R -2}$  that is 
like the propagator of the Shapiro-Virasoro model, but with only one set of
oscillators as in the generalized Veneziano model. In the following we
will rewrite it completely with the formalism of the Shapiro-Virasoro
model. This can be done   
by introducing the Pomeron propagator:
\begin{eqnarray}
\int_{0}^{\infty}  dt~  q^{2N-2} = \frac{2}{\pi \alpha'} {\hat{D}}~~;~~
{\hat{D}} \equiv \frac{\alpha'}{4 \pi} \int \frac{d^2 z}{|z|^2} z^{L_0 -1}
{\bar{z}}^{{\tilde{L}}_0 -1} ; |z| \equiv q =  e^{-\pi t}
\label{clopro}
\end{eqnarray}
and  rewriting the planar loop in the following compact form:
\begin{eqnarray}
\langle B_0 | {\hat{D}} | B_{M} \rangle~~;~~ 
| B_0 \rangle \equiv  \frac{T_{d-1}}{2} N
\prod_{n=1}^{\infty} e^{ a_{n}^{\dagger} \cdot
  {\tilde{a}}^{\dagger}_{n} } | p=0, 0_a , 0_{\tilde{a}} \rangle
\label{bound8}
\end{eqnarray}
where $| B_0 \rangle$ is the boundary state without any Reggeon on
it,
\begin{eqnarray}
T_{d-1} = \frac{\sqrt{\pi}}{2^{(d-10)/4} } ( 2 \pi \sqrt{\alpha'})^{-
  d/2 -1}
\label{Td}
\end{eqnarray}
and $ | B_{M} \rangle$ is instead the one with $M$ Reggeons given by:
\[
 | B_{M} \rangle =    N_{0}^{M} 
Tr ( \lambda^{a_1} \dots \lambda^{a_M} ) (2 \pi i )^M 
\int_{0}^{1} d \nu_{M} \int_{0}^{\nu_{M}} d
  \nu_{M-1} 
\dots \int_{0}^{\nu_{3}} d \nu_{2} \times  
\]
\begin{eqnarray}
\times \prod_{i=1}^{M} : e^{ i p_i \cdot Q ( e^{2
      i \pi \nu_i}) } :   | B_0 \rangle
\label{BM}
\end{eqnarray}
We want to stress once more that the normal ordering in the previous
equation is defined by taking the zero modes in the same exponential. 
Both the boundary states and the propagator are now states of the
Shapiro-Virasoro model. This means that we have rewritten the one-loop
planar diagram, where the states of the generalized Veneziano model
circulate in the loop, as a tree diagram of the Shapiro-Virasoro model
involving two boundary states
and a propagator.  This is what nowadays is
called open/closed string duality.
 
Besides the one-loop planar diagram in Eq. (\ref{onelo}), that is 
nowadays called the annulus diagram, also the non-planar 
and the non-orientable diagrams 
were constructed and studied. In particular the non-planar one, that 
is obtained as the planar one in Eq. (\ref{onelo}) but 
with two propagators multiplied with the twist operator
\begin{eqnarray}
\Omega =  e^{L_{-1}} (-1)^{R}~,
\label{twist}
\end{eqnarray}
 had unitarity violating
cuts that disappeared~\cite{LOVE1} if the dimension of the 
space-time $d=26$, leaving behind additional pole singularities. The
explicit form of the non-planar loop can be obtained following the
same steps done for the planar loop. One
gets for the non-planar loop the following amplitude:
\begin{eqnarray}
\langle B_{R} | {\hat{D}} | B_{M} \rangle
\label{nonpla}
\end{eqnarray}
where now both boundary states contain, respectively, $R$ and $M$
Reggeon states. 
The additional poles found in the non-planar loop 
were called Pomerons because they occur in the Pomeron 
sector, that today is called the closed string channel, to distinguish 
them from the Reggeons that instead occur in the 
Reggeon sector, that today is called the open string sector of the
planar and non-planar loop diagrams.  At that time in fact, the states
of the generalized Veneziano models were called Reggeons, while the
additional ones appearing in the non-planar loop were called
Pomerons.  The Reggeons correspond 
nowadays to open string states, while the Pomerons 
to closed string states.   These things are obvious now, but at that
time it took a while to show that the additional states appearing in
the Pomeron sector have to be  identified with those of the Shapiro-Virasoro
model. The proof that the spectrum was the same came rather early. This
was obtained by factorizing the
non-planar diagram  in the Pomeron channel~\cite{POME} as we have done
in Eq. (\ref{nonpla}). It was found
that  the states of the Pomeron channel  lie on  a linear Regge 
trajectory that 
has double intercept and half slope of the one of the Reggeons.
This follows immediately from the propagator ${\hat{D}}$ in 
Eq. (\ref{clopro}) that has
poles for values of the momentum of the Pomeron exchanged given by:
\begin{eqnarray}
2 - \frac{\alpha'}{2} p^2 = 2n
\label{clostri7}
\end{eqnarray}
that are exactly the values of the masses of the states of the
Shapiro-Virasoro model~\cite{OLISCHE},
while the Reggeon propagator in Eq. (\ref{proab}) 
has poles for values of momentum equal
to:
\begin{eqnarray}
1 - \alpha' p^2 = n
\label{opens67}
\end{eqnarray}
However, it was still not clear
that the Pomeron states interact among themselves as the states
of the Shapiro-Virasoro model. To show this  it was first  necessary
to construct  tree amplitudes containing both states 
of the generalized Veneziano model and of the 
Shapiro-Virasoro model~\cite{MIX}. They reduced to the amplitudes of
the generalized Veneziano (Shapiro-Virasoro) model if we have 
only external states of the generalized Veneziano (Shapiro-Virasoro) model.
Those amplitudes are called today disk amplitudes
containing both open and closed string states.  
They were constructed~\cite{MIX} 
by using for the Reggeon states  the vertex operators that we have 
discussed in Sect. (\ref{phys})  involving one set of harmonic oscillators
and for the Pomeron states the vertex operators  given in 
Eq. (\ref{veclo67}) that we rewrite here:
\begin{eqnarray}
V_{\alpha , \beta} ( z, {\bar{z}}, p ) = V_{\alpha} ( z, \frac{p}{2} ) 
V_{\beta} ( {\bar{z}}, \frac{p}{2} ) 
\label{veclo67b}
\end{eqnarray}
because now both component vertices contain the same set of
harmonic oscillators as in the generalized Veneziano model. 
Furthermore, each of the two vertices is separately normal 
ordered, but their product is nor normal ordered.  The amplitude
involving both kinds of states is then constructed by  taking 
the product of all vertices between the projective invariant vacuum and
integrating the Reggeons on the real axis in an ordered way and
the Pomerons in the upper half plane, as one does for a disk
amplitude.  

We have mentioned above that the two vertices are separately
normal ordered, but their product is not normal ordered.  
When we normal order
them we get, for instance for the tachyon of the Pomeron sector, a factor
$(z - {\bar{z}} )^{\alpha ' p^2 /2}$ that describes 
the Reggeon-Pomeron transition. This implies a direct 
coupling~\cite{CRESCHE} between
the $U(1)$  part of gauge field and the two-index 
antisymmetric field $B_{\mu \nu}$, called Kalb-Ramond field~\cite{KR},  
of the Pomeron sector, that makes the gauge field 
massive~\cite{CRESCHE}. 

It was then shown that, by factorizing 
the non-planal loop in the  Pomeron channel, one reproduced  
the scattering amplitude containing one state of the Shapiro-Virasoro
and a number of states of the generalized Veneziano model~\cite{REGPOM}. 
If we have also
external states belonging to the generalized Shapiro-Virasoro model, then by
factorizing the non-planar one loop amplitude in the pure Pomeron channel, 
one would obtain the tree amplitudes of the 
Shapiro-Virasoro model~\cite{REGPOM}. 

All this 
implies that the generalized Veneziano model and the 
Shapiro-Virasoro model  are not two independent models, but they 
are part of the same and unique model. In fact, if one 
started with the generalized
Veneziano model and added loop diagrams to implement unitarity, one found 
the appearence in the non-planar loop of additional 
states that had the same mass
and interaction of those of the Shapiro-Virasoro model.

The planar diagram,  written in Eq. (\ref{clo84}) in the closed string 
channel, is   divergent for large values of $t$. This divergence was
recognized to be due to  exchange, in the Pomeron channel, of 
the tachyon of the Shapiro-Virasoro model 
and of the dilaton~\cite{ADE2}. They correspond, respectively, to the
first two terms of the expansion:
\begin{eqnarray}
[ f_{1} (q)]^{-24} = e^{2 \pi t} + 24 + O \left( e^{-2\pi t} \right) 
\label{exp83}
\end{eqnarray}
The first one could be cancelled by an analytic continuation, while
the second one could be eliminated through a 
renormalization of the slope of the Regge
trajectory $\alpha'$~\cite{ADE2}.

We conclude  the discussion of the one-loop diagrams  by mentioning 
that the one-loop
diagram for the Shapiro-Virasoro model was computed 
by Shapiro~\cite{JS2} who also
found that the integrand was modular invariant.

The computation of  multiloop diagrams  requires a more advanced 
technology that was also
developed in the early days of the dual resonance model  few years
 before
the discovery of its connection to string theory. 
In order to compute  multiloop diagrams one needs first  to construct  
an object
that was called the $N$-Reggeon vertex and 
that has the properties of containing $N$
sets of harmonic oscillators, one for 
each external leg, and is such that, when we 
saturate it with $N$ physical states, we get  the corresponding 
$N$-point amplitude. In the following we will  discuss  how to
determine the 
$N$-Reggeon vertex.   

The first step toward the $N$-Reggeon vertex is
the Sciuto-Della Selva- Saito~\cite{SDS} 
vertex that includes two sets 
of harmonic oscillators that we denote with the indices 1 and 2. 
It is equal to:
\begin{eqnarray}
V_{SDS} = {}_2 \langle x=0, 0 | 
: \exp \left(-  \frac{1}{2 \alpha'} 
\oint_{0} dz X_{2}' (z) \cdot X_{1} (1 -z) \right) :
\label{SDS}
\end{eqnarray}
where $X$ is the quantity that we have called $Q$ 
in Eq. (\ref{qfv}) and the prime 
denotes a derivative with respect to z. It satisfies the important property 
of giving the vertex operator $V_{\alpha} (z=1)$ 
of an arbitrary state $| \alpha \rangle$ when we saturate
it with the corresponding state:
\begin{eqnarray}
V_{SDS} | \alpha \rangle_2 = V_{\alpha} (z=1)
\label{ver94}
\end{eqnarray}
A shortcoming of this vertex is that it is not invariant under a
cyclic 
permutation of the three legs.  A cyclic symmetric vertex has been 
constructed by Caneschi, Schwimmer and Veneziano~\cite{CSV}  by 
inserting the twist operator in Eq. (\ref{twist}).
But the  $3$-Reggeon vertex is not enough if we want to compute an 
arbitrary multiloop amplitude. We must generalize it to an 
arbitrary number of external legs. Such a vertex, that can be obtained
from the one in Eq. (\ref{SDS}) with a very direct procedure, or 
that can also be obtained by sewing together three-Reggeon vertices,  
has been written in its final form by 
Lovelace~\cite{LOVE1b}~\footnote{See also Ref.~\cite{OLIVE2}.   
Earlier papers on the 
$N$-Reggeon can be found in Ref.s~\cite{nregge}.}. 
Here we do not  derive it, but we give 
directly its expression written in Ref.~\cite{LOVE1b}:
\begin{eqnarray}
V_{N,0}&=&\int \frac{\prod_{i=1}^{N}d z_{i}}{dV_{abc}\prod_{i=1}^{N}[ 
V_{i}'(0)]}    
\,\prod_{i=1}^{N}\,[{}_{i}\!\!<\! x=0, O_{a}|] 
\,\delta(\sum_{i=1}^{N} p_{i})\nonumber\\
&&\;\;\prod_{\stackrel{\scriptstyle i,j=1}{i \neq j}}^{N}
\exp\left[-\frac{1}{2}\sum_{n,m=0}^{\infty} \!a_{n}^{(i)} 
D_{nm}( \Gamma V_{i}^{-1} V_{j}) \,a_{m}^{(j)}\right]
\label{3}
\end{eqnarray}
where $a_{0}^{(i)} \equiv\alpha_{0}^{i} =  \sqrt{2 \alpha'} {\hat{p}}_{i}$ 
is the momentum of particle $i$
and  the infinite matrix:
\[
D_{nm}(\gamma)=\frac{1}{m!}\sqrt{\frac{m}{n}}\partial_{z}^{m}[ 
\gamma(z)]^{n}|_{z=0}~;~n,m=1..~:~ D_{00}(\gamma)= - 
\log|\frac{D}{\sqrt{AD - BC}}|
\]
\begin{eqnarray}
D_{n0}=\frac{1}{\sqrt{n}}(\frac{B}{D})^{n}~~;~~      
 D_{0n}=\frac{1}{\sqrt{n}}(-\frac{C}{D})^{n}~~;~~
\gamma(z)=\frac{Az+B}{Cz+D} 
\label{9}
\end{eqnarray}
is a "representation" of the projective group corresponding to the 
conformal weight $\Delta=0$, that satisfies the eqs.:
\begin{eqnarray}
D_{nm}(\gamma_{1}\gamma_{2})=\sum_{l=1}^{\infty}D_{nl}(\gamma_{1}) 
D_{lm}(\gamma_{2}) + 
D_{n0}(\gamma_{1})\delta_{0m}+D_{0m}(\gamma_{2})\delta_{n0}
\label{10}
\end{eqnarray}
and
\begin{eqnarray}
D_{nm}(\gamma) = D_{mn}( \Gamma \gamma^{-1} \Gamma) \hspace{2cm} 
\Gamma(z)=\frac{1}{z}\label{11}
\end{eqnarray}
Finally $V_i$ is a projective transformation that maps $0, 1$ 
and $\infty$ into $z_{i-1},  z_i$ and $z_{i+1}$.

The previous vertex can be written in a more elegant form as follows:
\[
V_{N,0}=\int \frac{\prod_{i=1}^{N}d z_{i}}{dV_{abc}\prod_{i=1}^{N}[ 
V_{i}'(0)]}    
\,\prod_{i=1}^{N}\,[{}_{i}\!\!<\! x=0, O_{a}|] 
\,\delta(\sum_{i=1}^{N} p_{i})
\]
\[
 \exp \left\{\frac{i}{4\alpha'} 
\oint dz \partial X^{(i)}(z) {\hat{p}}_{i} \log
V_i'(z) \right\}
\]
\begin{eqnarray}
\exp\left\{-\frac{1}{2}
{\sum_{\stackrel{i,j=1}{i\neq j}}^{N}}\oint dz
\oint dy \partial X^{(i)}(z) \log[V_i(z) - V_j(y)] \partial
X^{(j)}(y)\right\}
\label{3bis}
\end{eqnarray}
where the quantities $X^{(i)}$ are what we called $Q$, namely 
the Fubini-Veneziano
field,  in the previous sections.
The $N$-Reggeon vertex that satisfies the important property of giving the
scattering amplitude of $N$ physical particle  when we saturate it with
their corresponding states, is the fundamental object for computing the
multiloop amplitudes. In fact, if we want to compute a $M$-loop amplitude
with $N$ external states, we need to start from the $(N+ 2M)$-Reggeon 
vertex and then we have to sew the M pairs together after having inserted a
propagator D. In this way we obtain
an amplitude that is not only integrated over the punctures 
$z_i \,\,(i=1 \dots N)$ of the $N$ external states, but also over the additional
$3h -3$ moduli  corresponding to the punctures variables of
 the states that we sew together and the integration 
 variable of the $M$ propagators.  $h$ is the number of loops.
The multiloop amplitudes have been obtained in this way already 
in 1970~\cite{LOVE3,ALE,AA} and, through the sewing procedure,
one obtained  
functions, as the period matrix, the abelian differentials, the prime form,
 etc., that
are well defined on Riemann surface! The only thing that was 
missing, was the correct measure of integrations over the $3h -3$
variables because it was technically not possible to let only 
the physical states to circulate in the loops.
This problem was solved  
only much later~\cite{DFLS, MANDE} when a BRST invariant 
formulation of string theory
 and the light-cone functional integral 
could be used for computing multiloops. They are two very 
different approaches that, however, gave the same result.
For the sake of completeness we write here the  planar $h$-loop
amplitude involving $M$ tachyons:
\begin{eqnarray}
A^{(h)}_M (p_1,\ldots,p_M) & = & N^h\,{\rm Tr}(\lambda^{a_1}
\cdots \lambda^{a_M})~
C_h\,\left[2 g_s \left( 2 \alpha ' \right)^{(d-2)/4} \right]^M    
\nonumber \\
& \times & \int [dm]^M_h\,
\prod_{i<j} \left[{{\exp\left({\cal G}^{(h)}(z_i,z_j)\right)}
\over{\sqrt{V'_i(0)\,V'_j(0)}}}\right]^{2\alpha ' p_i\cdot p_j}~~,
\label{hmastac}
\end{eqnarray}
where $N^h\,{\rm Tr}(\lambda^{a_1}
\cdots \lambda^{a_M})$ is the appropriate $U(N)$ Chan-Paton factor,
$g$ is the dimensionless open string coupling constant,
${\cal C}_h$ is a normalization factor given by
\begin{eqnarray}
C_h = {1\over{(2\pi)^{dh}}}~g_s^{2h-2}{1\over{(2\alpha ')^{d/2}}}~~,
\label{vertnorm}
\end{eqnarray}
and ${\cal G}^{(h)}$ is the $h$-loop bosonic Green function
\begin{eqnarray}
{\cal G}^{(h)}(z_i,z_j) = \log E^{(h)}(z_i,z_j) - {1\over 2} \int_{z_i}^{z_j}
\omega^\mu \, \left(2\pi {\rm Im}\tau_{\mu\nu}\right)^{-1}
\int_{z_i}^{z_j} \omega^\nu~~~,
\label{hgreen}
\end{eqnarray}
with $E^{(h)}(z_i,z_j)$ being the prime form,
$\omega^\mu$ ($\mu=1,\ldots, h$) the abelian
differentials and $\tau_{\mu\nu}$ the period matrix.
All these objects, as well as the measure on moduli space $[dm]^M_h$, can
be explicitly written in the Schottky parametrization of the Riemann
surface, and their expressions for arbitrary $h$ can be found for example
in Ref.~\cite{scho}. 
It is given by
\begin{eqnarray}
[dm]^M_h & = & \frac{1}{dV_{abc}} \prod_{i=1}^M \frac{dz_i}{ V_{i} ' (0)
 }
\prod_{\mu=1}^{h} \left[ \frac{dk_\mu \,d \xi_\mu \,d \eta_\mu}{k_\mu^2
\,(\xi_\mu - \eta_\mu)^2} ( 1- k_\mu )^2 \right]   \label{hmeasure} \\
& \times & \left[\det \left( - i \tau_{\mu \nu} \right) \right]^{-d/2}
{}~\prod_{\alpha}\,' \left[ \prod_{n=1}^{\infty} ( 1 - k_{\alpha}^{n})^{-d}
\prod_{n=2}^{\infty} ( 1 - k_{\alpha}^{n})^{2} \right]~~~.   \nonumber
\end{eqnarray}
where $k_{\mu}$ are the multipliers, $\xi_{\mu}$ and $\eta_{\mu}$ are the
fixed points of the generators of the Schottky group,

\section{From dual models to string theory}
\label{string}

The approach presented in the previous sections is a real bottom-up approach.
The experimental data were the driving force 
in the construction of the Veneziano 
model  and of its generalization to $N$ external legs. 
The rest of the work that we have described above consisted in 
deriving its properties. The result is, except for a tachyon,
a fully consistent quantum-relativistic model that was a source of  fascination
for those who worked in the field.   Although the model 
grew out of S-matrix theory
where the scattering amplitude is the 
only observable object, while the action or
the Lagrangian have not a central role, some people nevertheless started to 
investigate what was the underlying microscopic structure that gave 
rise to such a consistent and beautiful model. 
It turned out, as we know today, that this underlying
structure is that of a quantum-relativistic string.  However, the
process of connecting the dual resonance model (actually two of them 
the generalized Veneziano and the Shapiro-Virasoro model)
to  string theory
took several years from the original idea to a complete and convincing
proof of the conjecture. The original conjecture
was independently formulated by Nambu~\cite{NAMBU1,NAMBU2}, 
Nielsen~\cite{NIELSEN} and Susskind~\cite{LENNY}~\footnote{See 
also Ref.~\cite{other}.}. If we look at it in retrospective, it was 
at that time a fantastic idea that shows  the enormous physical
intuition of those who formulated it. On the other hand, it took
several years to digest it before one was able to derive from it all
the deep features of the dual resonance model. Because of this, the
idea that the underlying structure was that of a relativistic string,
did not really influence most of the research in the field up to 1973.
Let me try to explain why.
 
A common feature of the work of Ref.s~\cite{NAMBU1,NIELSEN,LENNY} 
is the suggestion that the infinite
number of oscillators, that one got through the factorization of the
dual resonance model, naturally comes out from a two-dimensional
free Lagrangian for the coordinate $X^{\mu} ( \tau , \sigma)$ 
of a one-dimensional string, that is
an obvious generalization of the Lagrangian that one writes for the
coordinate $X^{\mu} (\tau )$ of a 
pointlike object in the proper-time gauge:
\begin{eqnarray}
L \sim \frac{1}{2} \frac{d X}{d\tau} \cdot \frac{d X}{d\tau}
\Longrightarrow
L \sim  \frac{1}{2} \left[\frac{d X}{d\tau} \cdot \frac{d X}{d\tau} -
\frac{d X}{d\sigma} \cdot \frac{d X}{d\sigma} \right]
\label{strila}
\end{eqnarray}
Being this theory conformal invariant the Virasoro operators were also
constructed together with their algebra. In this very first
formulation, however, the Virasoro generators $L_n$ were just the
generators associated to the conformal symmetry of the string
world-sheet Lagrangian given in Eq. (\ref{strila}) as in any conformal
field theory. It was not clear at all
why they should imply the gauge conditions found
 by Virasoro or, in modern terms, why they should be zero classically.
The basic ingredient  to solve this problem was provided by 
 Nambu~\cite{NAMBU2} and Goto~\cite{GOTO}
who wrote the non-linear Lagrangian proportional to the area spanned   by
the string in the external target space. They proceeded in analogy
with the point particle and wrote the following action:
\begin{eqnarray}
\label{2.2.3}
S \sim \int \sqrt{ - d \sigma_{\mu \nu} d \sigma^{\mu \nu}}
\end{eqnarray}
where
\begin{equation}
\label{2.2.2}
d \sigma_{\mu \nu} = \frac{\partial X_{\mu}}{\partial \zeta^{\alpha}}
\frac{\partial X_{\nu}}{\partial \zeta^{\beta}} d \zeta^{\alpha} \wedge d 
\zeta^{\beta} = \frac{\partial X_{\mu}}{\partial \zeta^{\alpha}}
\frac{\partial X_{\nu}}{\partial \zeta^{\beta}} \epsilon^{\alpha \beta} d 
\sigma d \tau
\end{equation}
$X_{\mu}( \sigma, \tau)$ is the string   coordinate and   
$\zeta^{0}= \tau$ and $\zeta^{1} = \sigma$ are the coordinates of the  
string worldsheet.  $\epsilon^{\alpha \beta}$ is
an antisymmetric tensor with $\epsilon^{01} =1$.
Inserting eq. (\ref{2.2.2}) in (\ref{2.2.3}) and fixing the proportionality 
constant one gets the Nambu-Goto action~\cite{NAMBU2,GOTO}:
\begin{eqnarray}
\label{2.2.4}
S = - c T \int_{\tau_{i}}^{\tau_{f}} d \tau \int_{0}^{\pi} d \sigma 
\sqrt{ ( \dot{X} \cdot X' )^{2} - \dot{X}^{2} {X'}^{2}} 
\end{eqnarray}
where
\begin{eqnarray}
\label{2.2.5}
\dot{X}^{\mu} \equiv \frac{\partial X^{\mu}}{\partial \tau}
\hspace{2cm}
{X'}^{\mu} \equiv \frac{\partial X^{\mu}}{\partial \sigma}
\end{eqnarray}
and  $ T \equiv \frac{1}{2 \pi \alpha'}$  
is the string tension, that replaces the mass appearing in the
case of a point particle. In this formulation, the
string Lagrangian is invariant under any reparametrization of the
world-sheet coordinates $\sigma$ and $\tau$ and not only under the
conformal transformations. This, in fact, implies that the two-dimensional 
world-sheet energy-momentum tensor of the string is actually zero as
we will show later on.  But it took still a few years to connect
the Nambu-Goto action to the properties of the dual resonance model.
In the meantime an analogue
model was formulated~\cite{FN} that reproduced the tree and loop
amplitudes of the generalized Veneziano model. This approach
anticipated by several years the path integral derivation of dual
amplitudes.  It  was very
closely related to the functional integral formulation of
Ref.s~\cite{HSV}.

However, one needed to wait until 1973 with the paper of 
Goddard, Goldstone, Rebbi and Thorn~\cite{GGRT}, where 
the Nambu-Goto action was correctly treated, all its consequences
were derived and it became completely clear that the structure
underlying the dual resonance model was that of a quantum-relativistic
string. The equation of motion
for the string were derived from the action in Eq. (\ref{2.2.4}) by
imposing  $\delta S =0$ for variations such that 
$\delta X^{\mu} (\tau_{i}) = \delta X^{\mu}( \tau_{f}) =0$. One gets:
\begin{equation}
\label{2.2.10}
\delta S = \int_{\tau_{i}}^{\tau_{f}} \left[ \int_{0}^{\pi} d \sigma \left( 
- \frac{\partial}{\partial \tau} \frac{\partial L}{\partial 
\dot{X}^{\mu}} - \frac{\partial}{\partial \sigma} \frac{\partial L}{\partial 
{X'}^{\mu}} \right) \delta  X^{\mu} + \frac{\partial L}{ \partial {X'}^{\mu}} 
\delta X^{\mu} |_{\sigma=0}^{\sigma= \pi} \right] =0
\end{equation} 
where $L$ is the Lagrangian in Eq. (\ref{2.2.4}).
Since $\delta X^{\mu}$ is arbitrary, from eq. (\ref{2.2.10}) one gets 
the Euler-Lagrange 
equation of motion
\begin{eqnarray}
\label{2.2.11}
\frac{\partial}{\partial \tau} \frac{\partial L}{\partial 
\dot{X}^{\mu}} + \frac{\partial}{\partial \sigma} \frac{\partial L}{\partial 
{X'}^{\mu}} \equiv \frac{\partial}{\partial \zeta^{\alpha}} \left( 
\frac{\partial L}{ \partial( \frac{\partial X^{\mu}}{\partial 
\zeta^{\alpha}})}\right) =0
\end{eqnarray}
and the boundary conditions 
\begin{eqnarray}
\label{2.2.12}
\frac{\partial L}{ \partial {X'}^{\mu}}=0 \hspace{1cm} or \hspace{1cm}
\delta X_{\mu} =0
\hspace{1cm} at \hspace{1cm}
\sigma=0, \pi
\end{eqnarray}
for an open string and 
\begin{eqnarray}
\label{2.2.13}
X^{\mu}( \tau,0) = X^{\mu} (\tau, \pi)
\end{eqnarray}
for a closed string. In the case of an open string, the first kind
of boundary condition in Eq.(\ref{2.2.12}) corresponds to Neumann boundary 
conditions, while the second one to Dirichlet boundary conditions. Only
the Neumann boundary conditions preserve the translation invariance of
the theory and, therefore, they were mostly used in the early days of
string theory. It must be stressed, however, that Dirichlet 
boundary conditions were already discussed and used in the early days
of string theory for constructing models with off-shell
states~\cite{CF}. 

{From} Eq. (\ref{2.2.4}) one can compute the momentum density along
the string:
\begin{eqnarray}
\frac{ \partial L}{\partial \dot{X}^{\mu}} \equiv P_{\mu} = cT \frac{ 
\dot{X}_{\mu} {X'}^{2} - {X'}_{\mu} ( \dot{X} \cdot X')}{
\sqrt{ ( \dot{X} \cdot X' )^{2} - \dot{X}^{2} {X'}^{2}} }
\label{2.2.14}
\end{eqnarray} 
and obtain the following constraints between the dynamical variables
$X^{\mu}$ and $ P^{\mu}$:
\begin{eqnarray}
\label{2.2.20}
c^{2} T^{2} {x'}^{2} + P^{2} = x' \cdot P =0
\end{eqnarray}
They are a consequence of the reparametrization invariance of the string
Lagrangian. Because of this one can choose the orthonormal gauge
specified by the conditions:
\begin{eqnarray}
{\dot{X}}^2 + {X'}^2 = {\dot{X}} \cdot X' =0
\label{ortho}
\end{eqnarray}
that nowadays is called conformal gauge.
In this gauge eq. (\ref{2.2.14}) becomes:
\begin{eqnarray}
P_{\mu} = c T {\dot{X}}_{\mu}  \hspace{2cm} \frac{\partial L}{\partial 
{X'}^{\mu}} = - cT X_{\mu}'
\label{momeor}
\end{eqnarray}
and therefore the eq. of motion in eq.(\ref{2.2.11}) becomes:
\begin{eqnarray}
{\ddot{X}}_{\mu} - X_{\mu}'' =0
\label{orthoeq}
\end{eqnarray}
while the boundary condition in eq.(\ref{2.2.12}) becomes:
\begin{eqnarray}
X_{\mu} ' (\sigma =0, \pi ) =0
\label{orthobc}
\end{eqnarray}
The most general solution of the eq. of motion and of the boundary conditions
can be written as follows:
\begin{eqnarray}
\label{2.3.11}
X^{\mu} ( \tau, \sigma) = q^{\mu} + 2 \alpha' p^{\mu} \tau + i 
\sqrt{2 \alpha'}\sum_{n=1}^{\infty} 
[ a_{n}^{\mu} e^{-in \tau} - a_{n}^{+ \mu} e^{in\tau}] \frac{ cosn 
\sigma}{\sqrt{n}}
\end{eqnarray}
for an open string and
\[
X^{\mu} ( \tau, \sigma) = q^{\mu} + 2 \alpha' p^{\mu} \tau + \frac{i}{2}
 \sqrt{2 \alpha '} \sum_{n=1}^{\infty} 
[ {\tilde{a}}_{n}^{\mu} e^{- 2in (\tau + \sigma)} - 
{\tilde{a}}_{n}^{ + \mu } e^{2 in(\tau + 
\sigma)}] \frac{ 1}{\sqrt{n}} +
\]
\begin{eqnarray}
\label{2.3.12}
+ \frac{i}{2} \sqrt{2 \alpha'}\sum_{n=1}^{\infty} 
[ {{a}}_{n}^{\mu} e^{- 2in (\tau - \sigma)} - {{a}}_{n}^{ + \mu } 
e^{2in(\tau - \sigma)}] \frac{ 1}{\sqrt{n}}
\end{eqnarray}
for a closed string. This procedure really shows that, starting from
the Nambu-Goto action, one can choose a gauge (the orthonormal or
conformal gauge) where the equation of motion of the string becomes
the two-dimensional D'Alembert equation in
Eq. (\ref{orthoeq}). Furthermore, the invariance under
reparametrization of the Nambu-Goto action implies that the
two-dimensional energy-momentum tensor is identically zero at the
classical level (See Eq. (\ref{ortho})). 

As the Lorentz gauge in QED the orthonormal gauge does not fix
completely the gauge. We can still perform reparametrizations that leave in
the conformal gauge: they are  conformal transformatiuons.  
Introducing the variable $ z = e^{i \tau}$ the generators of the 
conformal transformations for the open string can be written as
follows:
\begin{eqnarray}
\label{2.3.27}
L_{n} = \frac{1}{2 \pi i} \oint dz z^{n+1} \left[ - \frac{1}{4 \alpha'}
\left( \frac{\partial X^{\mu}}{\partial z} \right)^2 \right] =  
\frac{1}{2} \sum_{m=-\infty}^{\infty} \alpha_{n-m} \cdot 
\alpha_{m} =0
\end{eqnarray}
where
\begin{eqnarray}
\label{2.3.26}
\alpha_{n}^{\mu} = \left\{ \begin{array}{cc}
              \sqrt{n} a_{n}^{\mu}  & if \,\,\, n>0 \\
                  \sqrt{2 \alpha'}    p^{\mu}       & if \,\,\, n=0  \\
              \sqrt{n} a_{n}^{\dagger \mu} & if \,\,\, n<0
              \end{array} \right.
\end{eqnarray}
They are zero as a consequence of Eq.s (\ref{2.2.20}) that in the conformal
gauge become Eq.s (\ref{ortho}). In the case of a closed string we get
instead:
\begin{eqnarray}
\label{2.3.33}
{\tilde L}_{n} = \frac{1}{2 \pi i} \oint dz z^{n+1} \left[ - \frac{1}{\alpha'}
\left( \frac{\partial X^{\mu}}{\partial z} \right)^{2} \right] =0
\end{eqnarray}
\begin{eqnarray}
\label{2.3.34}
{{L}}_{n} = \frac{1}{2 \pi i} \oint d {\bar{z}} {\bar{z}}^{n+1} \left[ - 
\frac{1}{\alpha'} \left( \frac{\partial X^{\mu}}{\partial {\bar{z}}} 
\right)^{2} \right] =0
\end{eqnarray}
In terms of the harmonic oscillators
introduced in eq. (\ref{2.3.12}) we get
\begin{eqnarray}
\label{2.3.35}
L_{n} = \frac{1}{2} \sum_{m= - 
\infty}^{\infty} \alpha_{m} \cdot \alpha_{n-m} =0 
\hspace{.5cm}; \hspace{.5cm}
{\tilde{L}}_{n} = \frac{1}{2} \sum_{m= - \infty}^{\infty}
{\tilde{\alpha}}_{m} 
\cdot 
{\tilde{\alpha}}_{n-m} =0
\end{eqnarray}
where for the non-zero modes we have used the convention in 
(\ref{2.3.26}), while the zero mode is given by:
\begin{eqnarray}
\label{2.3.36}
\alpha _{0}^{\mu} = {\tilde{\alpha}}_{0}^{\mu} = 
\sqrt{2 \alpha'}\frac{p^{\mu}}{2}
\end{eqnarray}
In conclusion, the fact that we have reparametrization invariance 
 implies that the Virasoro generators are classically
identically zero. When we quantize the theory one cannot and also
does not need to  impose that they are vanishing  at the operator
level. They are imposed as
conditions characterizing the  physical states.
\begin{eqnarray}
\label{2.6.10}
\langle  Phys '| L_{n} | Phys \rangle = \langle Phys ' | (L_{0} -1 ) 
|Phys \rangle =0~~;~~   n \neq 0
\end{eqnarray}
These equations are satisfied if we require:
\begin{eqnarray}
\label{2.6.9}
L_{n} | Phys> = ( L_{0} -1) | Phys> =0
\end{eqnarray}
The extra factor $-1$ in the previous equations comes from the normal
ordering as explained in Eq. (\ref{zeropo1}).

The authors of Ref.~\cite{GGRT} further specified the gauge by fixing
it completely. They introduced  the light-cone gauge specified by imposing the
condition:
\begin{eqnarray}
\label{2.4.1}
X^{+} = 2 \alpha ' p^{+} \tau
\end{eqnarray}
where
\begin{eqnarray}
\label{2.4.2}
X^{\pm} = \frac{ X^{0} \pm X^{d-1} }{\sqrt{2}} \hspace{2cm}
X_{\pm} = \frac{X_0 \pm X_{d-1}}{\sqrt{2}}
\end{eqnarray}
In this gauge the only physical degrees of freedom are the transverse
ones. In fact the components along the directions $0$ and $d-1$ can be
expressed in terms of the transverse ones by inserting
Eq. (\ref{2.4.1}) in the constraints in Eq. (\ref{ortho}) and getting:
\begin{eqnarray}
\label{2.4.5}
\dot{X}^{-} = \frac{1}{4 \alpha' p^{+}} ( \dot{X}_{i}^{2} + {X'}_{i}^{2})
\hspace{1cm}
{X'}^{-} = \frac{1}{2 \alpha' p^{+}} \dot{X}_{i} \cdot {X'}_{i}
\end{eqnarray}
that up to a constant of integration determine completely $X^{-}$  as a
function of $X^{i}$. In terms of oscillators we get
\begin{eqnarray}
\label{2.4.9}
\alpha_{n}^{+} = 0~~~;~~~ 
\sqrt{2 \alpha'} 
\alpha_{n}^{-} = \frac{1}{2 p^{+}} \sum_{m = - \infty}^{\infty} 
\alpha_{n-m}^{i} 
\alpha_{m}^{i}
\hspace{1cm} n \neq 0
\end{eqnarray}
for an open string and
\begin{eqnarray}
\label{2.4.10}
\alpha_{n}^{+} = {\tilde{\alpha}}_{n}^{+} = 0 \hspace{2cm} n \neq 0
\end{eqnarray} 
together with
\[
\sqrt{2 \alpha'} 
\alpha_{n}^{-} = \frac{1}{2 p^{+}} \sum_{m = - \infty}^{\infty} 
\alpha_{n-m}^{i} 
\alpha_{m}^{i}
\]
\begin{equation}
\label{2.4.12}
\sqrt{2 \alpha'} 
{\tilde{\alpha}}_{n}^{-} = \frac{1}{2 p^{+}} \sum_{m = - \infty}^{\infty} 
{\tilde{\alpha}}_{n-m}^{i} {\tilde{\alpha}}_{m}^{i}
\end{equation}
in the case of a closed string.
 
This shows that the physical states are described only by the
transverse oscillators having  only  $d-2$ components. 
Those transverse oscillators correspond to
the transverse DDF operators that we have discussed in Section~\ref{DDF}. 
The authors of Ref.~\cite{GGRT} also constructed  the  Lorentz generators
only in terms of the transverse oscillators and they  showed that they
satisfy the correct Lorentz algebra only if the space-time dimension
is $d=26$. In this way the spectrum of the dual resonance model was
completely reproduced starting from the Nambu-Goto action if $d=26$!
On the other hand, the choice of $d=26$ is a necessity if we 
want to keep Lorentz invariance!

Immediately after this, the interaction was also included 
either by adding a term describing the
interaction of the string with an external gauge field~\cite{SUPGRU}
or by using  a  functional formalism~\cite{SM1,GS2}.
 
In the following we will give some detail only of the first approach
for the case of an open string. A way to describe the string
interaction is by adding to the free string action an additional term
that describes the interaction of the string with an external field. 
\begin{eqnarray}
\label{3.1.1}
S_{INT} = \int d^{D} y \Phi_{L} (y) J_{L} (y)
\end{eqnarray}
where $\Phi_{L} (y)$ is the external field and $J_{L}$ is the current generated
by the string. The index $L$ stands for possible Lorentz indices that are
saturated in order to have a Lorentz invariant action.

In the case of a point particle,  such an interaction 
term will not give  any information on the self-interaction of a particle.

In the case of a string, instead, we will see that 
$S_{INT}$  will describe the interaction among
strings because the  external fields that can consistently interact with
a string are only those that correspond to the various states of the string,
as it will become clear in the discussion below.

This is a consequence of the fact that, for the sake of consistency, we must
put the following restrictions on $S_{INT}$:
\begin{itemize}
\item{It must be a well defined operator in the space spanned by the
string oscillators.}            
\item{It must preserve the invariances of the free string theory. In
particular, in the "conformal gauge" it must be conformal invariant.}
\item{ In the case of an open string, the interaction occurs at the end
point of a string (say at $\sigma=0$). This follows from the fact that two open
strings interact attaching to each other at the end points.}
\end {itemize}      
The simplest scalar current generated by the motion of a string can be written
as follows
\begin{eqnarray}
\label{3.1.2}
J(y) = \int d \tau \int d \sigma \delta (\sigma) \delta^{(d)} [ y^{\mu} - 
x^{\mu}(\tau, \sigma)]
\end{eqnarray}
where $\delta (\sigma)$ has been introduced 
because the interaction occurs at the 
end of the string. For the sake of simplicity we omit to write a coupling 
constant $g$  in (\ref{3.1.2}).

Inserting (\ref{3.1.2}) in (\ref{3.1.1}) and using for the scalar external 
field $\Phi (y) = e^{i 
k \cdot y}$  a plane wave, we get the following interaction:
\begin{eqnarray}
\label{3.1.3}
S_{INT} = \int d \tau : e^{i k \cdot X ( \tau, 0)} :
\end{eqnarray}
where the normal ordering has been introduced in order to have a well defined
operator.
The invariance of (\ref{3.1.3}) under a conformal transformation $\tau 
\rightarrow w ( \tau)$  requires the following identity:
\begin{eqnarray}
\label{3.1.4}
S_{INT} = \int d \tau : e^{i k \cdot X(\tau,0)} : \,\, = \int d w : e^{ik 
\cdot X(w,0)}:
\end{eqnarray}
or, in other words, that
\begin{eqnarray}
\label{3.1.5}
: e^{i k \cdot X(\tau,0)} : \Longrightarrow w' ( \tau) : e^{i k \cdot X(w,0)} :
\end{eqnarray}
This means that the integrand in Eq. (\ref{3.1.4}) must be a conformal
field with conformal dimension equal to one  and this happens only if
$\alpha' k^2 =1$. The external field corresponds then  to the 
tachyonic lowest state of the open  string. 
Another simple current generated by the string is given by:
\begin{eqnarray}
\label{3.1.14}
J_{\mu} (y) = \int d \tau \int d \sigma \delta ( \sigma) \dot{X}_{\mu} ( 
\tau, \sigma) \delta^{(d)} ( y - X ( \tau, \sigma) )
\end{eqnarray}
Inserting (\ref{3.1.14}) in (\ref{3.1.1}) we get
\begin{equation}
\label{3.1.15}
S_{INT} = \int d \tau \dot{X}_{\mu}( \tau, 0) \epsilon^{\mu} e^{ i k \cdot 
X(\tau,0)}
\end{equation}
if we use a plane wave for $\Phi_{\mu}(y) = \epsilon_{\mu} e^{i k \cdot y}$. 
The vertex operator in 
eq. (\ref{3.1.15}) is conformal invariant only if
\begin{equation}
\label{3.1.19}
k^{2} = \epsilon \cdot k =0
\end{equation}
and, therefore, the external vector must be the massless photon state of the
string. We can generalize this procedure to an arbitrary external
field and the result is that we can only use external fields that
correspond to on shell physical states of the string.

This procedure has been 
extended in Ref.~\cite{SUPGRU} to the case of external gravitons by 
introducing in the Nambu-Goto action a target space metric and 
obtaining the vertex operator for the graviton that is a massless state
in the closed string theory.  Remember that, at that time, this could 
have been done only
with the Nambu-Goto action because  the $\sigma$-model 
action was introduced  only in 1976 first for the
point particle~\cite{POINT} and then for the string~\cite{STRING}. 
As in the case of the photon it turned out that the external
field corresponding to the graviton was required to be on shell.
This condition is the precursor of the equations of motion 
that one obtains from the $\sigma$-model action requiring 
the vanishing of the $\beta$-function~\cite{FRIEDAN}. 

One can then compute the probability amplitude for the emission of a
number of string states corresponding to the various external fields, from an
initial string state to a final one. This amplitude gives precisely
the $N$-point amplitude that we discussed in the previous 
sections~\cite{SUPGRU}. In
particular, one learns that, in the case of the open string, the
Fubini-Veneziano field is just the string coordinate computed at
$\sigma =0$:
\begin{eqnarray}
Q^{\mu} (z) \equiv X^{\mu} ( z, \sigma =0)~~;~~ z = e^{i \tau} 
\label{ide1}
\end{eqnarray} 
In the case of a closed string we get instead:
\begin{eqnarray}
Q^{\mu} (z, {\bar{z}}) \equiv X^{\mu} ( z, {\bar{z}})~~;~~ z = e^{2 i
  (\tau - \sigma)}~,~ {\bar{z}} = e^{2i(\tau + \sigma)} 
\label{ide2}
\end{eqnarray} 

Finally, let me mention that with the functional approach
Mandelstam~\cite{SM1} and Cremmer and Gervais~\cite{CG}  computed the
interaction between three arbitrary physical string states and reproduced
in this way the coupling of three DDF states given in Eq. (\ref{ver52b}) and
obtained in  Ref.~\cite{ADDF} by  using the operator formalism. 
At this point it was completely  clear that the structure underlying the
generalized Veneziano model was that of an open relativistic string,
while that underlying   the Shapiro-Virasoro model was that of a
closed relativistic string. Furthermore,  these two theories are not
independent because, if one starts from an open string theory, one gets
automatically closed strings by loop corrections. 

\section{Conclusions}
\label{conclu}

In this contribution, we have gone through the developments that led
from the construction of the dual resonance model to the bosonic 
string theory trying
as much as possible to include all the necessary technical details.  
This is because we believe that they are not only important from
an historical point of view, but are also still part of the formalism that
one uses today in many string calculations.  We have tried to be 
as complete and objective  as possible, but  it could very well be that
some of those who participated in the research of  these years, will
not agree with some or even many of the statements we made. 
We apologize to those we 
have forgotten to mention or we have not mentioned 
as they would have liked.

Finally, after having gone through the developments  of these years,
my thoughts go to Sergio Fubini who shared with me and Gabriele many of
the ideas described here and who is deeply missed, and to my friends
from Florence, Naples and Turin for a pleasant collaboration in many
papers discussed here.

\section*{Acknowledgments}   

I thank R. Marotta and I. Pesando for a critical reading of the manuscript.

\printindex
\end{document}